\documentclass[aps,prd,groupedaddress]{revtex4}
\usepackage{graphicx}
\usepackage{graphpap}
\usepackage{dcolumn}
\usepackage{bm}

\begin{document}


\noindent DESY 11-099\\
\noindent LPSC 11-128\\
\noindent MS-TP-11-12\\
\vspace*{5mm}

\title{Impact of squark flavour violation on neutralino dark matter}

\author{Bj\"orn Herrmann}
\email[]{bjoern.herrmann@desy.de}
\affiliation{Deutsches Elektronen-Synchrotron (DESY), Notkestra{\ss}e 85, D-22603 Hamburg, Germany}

\author{Michael Klasen}
\email[]{michael.klasen@uni-muenster.de}
\affiliation{Institut f\"ur Theoretische Physik,
 Universit\"at M\"unster,
 Wilhelm-Klemm-Stra{\ss}e 9, D-48149 M\"unster, Germany}

\author{Quentin Le Boulc'h}
\email[]{leboulch@lpsc.in2p3.fr}
\affiliation{Laboratoire de Physique Subatomique et de Cosmologie,
 Universit\'e Joseph Fourier/CNRS-IN2P3/INPG,
 53 Avenue des Martyrs, F-38026 Grenoble, France}

\date{\today}

\begin{abstract}
We discuss the possibility of new sources of flavour violation in the squark sector of supersymmetric models in the context of the dark matter relic density. We show that the corresponding non-minimal flavour violation terms in the squark mass matrices can have an important impact on the thermally averaged (co)annihilation cross section of the neutralino, and in consequence can modify its predicted relic density. We discuss in detail the relevant effects and present a numerical study of neutralino annihilation and coannihilation in this context. We also comment on the LHC phenomenology of the corresponding scenarios.
\end{abstract}


\maketitle

\section{Introduction \label{sec1}} 

Among the numerous extensions of the standard model of particle physics, supersymmetry ranks among the most popular ones. In particular, the Minimal Supersymmetric Standard Model (MSSM) is probably the best studied scenario of new physics. It allows to cure the hierarchy problem by stabilizing the Higgs mass and leads to gauge coupling unification. Moreover, it includes promising candidates for dark matter, whose presence remains the most compelling observational evidence for physics beyond the standard model.

Nevertheless, several open questions remain, e.g., concerning the flavour structure of the theory. While models with minimal flavour violation (MFV) \cite{Hall, D'Ambrosio, Altmannshofer} assume that the mechanism of flavour violation is the same as in the standard model, the framework of non-minimal flavour violation (NMFV) allows for new sources of flavour mixing, depending on the exact mechanism of supersymmetry breaking. In the former case, the rotation of the Yukawa couplings from gauge to mass eigenstates remains the only source of flavour violation, and thus all flavour-violating interactions are parameterized through the CKM- and PMNS-matrices as in the standard model. For NMFV, the terms originating from the additional sources are not related to these matrices, such that they are considered as additional parameters at the SUSY scale.

In recent years, supersymmetric scenarios beyond minimal flavour violation have received considerable attention in the community, especially in the context of signatures at current or future colliders. Concerning (s)quark flavour violation, the production and subsequent decays of squarks and gluinos at the Large Hadron Collider (LHC) have been studied, e.g., in Refs.\ \cite{NMFV_mSUGRA, NMFV_GMSB, HurthPorod, WienGluino, NMFV_Squark1, NMFV_Squark2}. Apart from the production of superpartners at colliders, the flavour-violating terms also appear in the (co)annihilation cross section of the neutralino, which is needed in the calculation of its relic density for a given scenario. 

In case of neutralino pair annihilation into quarks, the squarks appear as internal propagators. Additional flavour-violating terms can then increase the relative contributions of these diagrams since the mass splitting of the squarks is modified. Moreover, NMFV allows for efficient annihilation into final states, that are forbidden in the case of MFV. Flavour violating effects are also important in the case of coannihilations of a neutralino with a squark, since the latter is then an external particle. The importance of such processes crucially depends on the mass difference of neutralino and squark. The increased mass splitting of the squarks can therefore have an important impact on coannihilation processes. Finally, also in this case new final states are opened, leading to additional coannihilation channels. Recently, the impact of non-minimal flavour violation in the sector of sleptons on the coannihilation of a neutralino with a slepton has been discussed in Ref.\ \cite{LFV_CoAnn}.

The aim of the present paper is to provide a study of quark flavour violation in the context of neutralino dark matter. In this context, possible flavour-mixing effects are generally not considered in the literature. We present a detailed analysis of neutralino pair annihilation and neutralino-squark coannihilation in the MSSM beyond MFV. In Sec.\ \ref{sec2}, we will briefly introduce the MSSM with NMFV in the sector of squarks and discuss its parameterization. The role of generation mixing in the context of neutralino (co)annihilations is discussed in detail in Sec.\ \ref{sec3}. Sec.\ \ref{sec4} is then devoted to numerical examples in the context of neutralino (co)annihilation and its relic density. A discussion of LHC phenomenology for the corresponding scenarios follows in Sec.\ \ref{sec5}. Finally, conclusions are given in Sec.\ \ref{sec6}.

\section{The MSSM beyond minimal flavour violation \label{sec2}}

In the standard model, the only source of flavour violation are the Yukawa interactions, since their diagonalization leads to a mismatch between flavour and mass eigenstates of quarks and leptons. The flavour structure of the quark sector is very well described by the Cabibbo-Kobayashi-Maskawa (CKM) matrix, which only appears in charged currents, while flavour changing neutral currents are strongly suppressed. In supersymmetric theories with minimal flavour violation (MFV), the Yukawa matrices remain the only source of flavour violation, so that all flavour violating interactions of squarks are also related to the CKM-matrix. However, new sources of flavour violation may be present in supersymmetric models, especially if they are embedded in a grand unification framework. Depending on the exact realization and the involved representations, specific relations to the Yukawa matrices can lead to flavour non-diagonal entries in the soft-breaking terms. These are not related to the CKM-matrix and the corresponding framework is in consequence referred to as non-minimal flavour violation (NMFV). 

Considering the most general flavour structure, the squark mass matrices at the electroweak scale take the form
\begin{equation}
	{\cal M}^2_{\tilde{q}} ~=~ \left( \begin{array}{cc} {\cal M}^2_{\tilde{q},\rm LL} & {\cal M}^2_{\tilde{q},\rm LR} \\[2mm] {\cal M}^2_{\tilde{q},\rm RL} & {\cal M}^2_{\tilde{q},\rm RR} \end{array} \right)
\label{eq:massmatrix}
\end{equation}
for $q=u,d$, respectively. Their diagonal blocks are given by
\begin{eqnarray}
	{\cal M}^2_{\tilde{d},{\rm RR}} &=& M^2_{\tilde{D}} + m^2_d + e_d m_Z^2 \sin^2\theta_W \cos 2\beta, \label{eq:RR}\\
	{\cal M}^2_{\tilde{d},{\rm LL}} &=& M^2_{\tilde{Q}} + m^2_d + m_Z^2 \cos 2\beta (I_d-e_d \sin^2\theta_W),\\
	{\cal M}^2_{\tilde{u},{\rm RR}} &=& M^2_{\tilde{U}} + m^2_u + e_u m_Z^2 \sin^2\theta_W \cos 2\beta,\\
	{\cal M}^2_{\tilde{u},{\rm LL}} &=& V_{\rm CKM} M^2_{\tilde{Q}} V_{\rm CKM}^{\dag} + m^2_u + m_Z^2 \cos 2\beta (I_u-e_u \sin^2\theta_W),
\end{eqnarray}
where $M_{\tilde{Q}}$, $M_{\tilde{U}}$, and $M_{\tilde{D}}$ are the soft-breaking mass terms of the squarks. The diagonal mass matrices of up- and down-type quarks are denoted $m_u$ and $m_d$. Due to the SU(2) symmetry, the left-left entries are related through the CKM-matrix $V_{\rm CKM}$. The above expressions also involve the mass $m_Z$ of the Z-boson, the fractional electric charge $e_q$ and the weak isospin $I_q$ of the (s)quark, the weak mixing angle $\theta_W$, and the Higgs-mixing parameter $\beta$ defined through the ratio of the vacuum expectation values of the two Higgs doublets, $\tan\beta=v_u/v_d$.

The off-diagonal blocks of the matrix in Eq.\ (\ref{eq:massmatrix}) are given by
\begin{eqnarray}
	{\cal M}^2_{\tilde{u},{\rm RL}} ~=~ \big({\cal M}^2_{\tilde{u},{\rm LR}}\big)^{\dag} &=& \frac{v_u}{\sqrt{2}} T_U - \mu^* m_u \cot\beta ,\label{eq:RLu} \\
	{\cal M}^2_{\tilde{d},{\rm RL}} ~=~ \big({\cal M}^2_{\tilde{d},{\rm LR}}\big)^{\dag} &=& \frac{v_d}{\sqrt{2}} T_D - \mu^* m_d \tan\beta , \label{eq:RLd}
\end{eqnarray}
where $\mu$ is the Higgs mass parameter. The trilinear matrices $T_{U,D}$ are related to the soft-breaking matrices $A_{u,d}$ and the respective Yukawa matrices $Y_{u,d}$ through $\left( T_{U,D} \right)_{ij} = \left( A_{u,d} \right)_{ij} \left( Y_{u,d} \right)_{ij}$. 
At the GUT scale, the usual CMSSM condition $\left( A_u \right)_{33} = \left( A_d \right)_{33} = A_0$ applies, and the numerical values for $A_{u,d}$ at the SUSY scale are obtained through renormalization group running. All parameters appearing in Eqs.\ (\ref{eq:RR}) to (\ref{eq:RLd}) are understood to be in the super-CKM basis, where the neutral currents are flavour-diagonal and the quark (but not the squark) fields are in the mass eigenstate basis \cite{Gabbiani, Hagelin}.

In order to have a scenario-independent and dimensionless parameterization of flavour-mixing, the off-diagonal entries are usually normalized to the diagonal ones according to
\begin{eqnarray}
	\delta^{\rm LL}_{ij} &=& \big(M^2_{\tilde{Q}}\big)_{ij}\,/\,\sqrt{ \big(M^2_{\tilde{Q}}\big)_{ii} \big(M^2_{\tilde{Q}}\big)_{jj} }, \\ 			  
	\delta^{u,\rm RR}_{ij} &=& \big(M^2_{\tilde{U}}\big)_{ij}\,/\,\sqrt{ \big(M^2_{\tilde{U}}\big)_{ii} \big(M^2_{\tilde{U}}\big)_{jj} }, \\
	\delta^{d,\rm RR}_{ij} &=& \big(M^2_{\tilde{D}}\big)_{ij}\,/\,\sqrt{ \big(M^2_{\tilde{D}}\big)_{ii} \big(M^2_{\tilde{D}}\big)_{jj} }, \\
	\delta^{u,\rm RL}_{ij} &=& \frac{v_u}{\sqrt{2}}\big(T_U\big)_{ij}\,/\,\sqrt{ \big(M^2_{\tilde{Q}}\big)_{ii} \big(M^2_{\tilde{U}}\big)_{jj} }, \\
	\delta^{d,\rm RL}_{ij} &=& \frac{v_d}{\sqrt{2}}\big(T_D\big)_{ij}\,/\,\sqrt{ \big(M^2_{\tilde{Q}}\big)_{ii} \big(M^2_{\tilde{D}}\big)_{jj} }, \\
	\delta^{u,\rm LR}_{ij} &=& \frac{v_u}{\sqrt{2}}\big(T_U^{\dag}\big)_{ij}\,/\,\sqrt{ \big(M^2_{\tilde{U}}\big)_{ii} \big(M^2_{\tilde{Q}}\big)_{jj} }, \\
	\delta^{d,\rm LR}_{ij} &=& \frac{v_d}{\sqrt{2}}\big(T_D^{\dag}\big)_{ij}\,/\,\sqrt{ \big(M^2_{\tilde{D}}\big)_{ii}  \big(M^2_{\tilde{Q}}\big)_{jj} }.
\end{eqnarray}
The normalization factor is defined in terms of the corresponding diagonal elements of the soft-breaking matrices. We emphasize that the following numerical analysis is based on the diagonalisation of the full $6\times 6$ mass matrices. This is realized by introducing two rotation matrices, such that
\begin{equation}
	{\cal R}_{\tilde{q}} {\cal M}^2_{\tilde{q}} {\cal R}_{\tilde{q}}^{\dag} ~=~ {\rm diag}\left( m^2_{\tilde{q}_1}, \dots, m^2_{\tilde{q}_6} \right)
\label{eq:massdiag}
\end{equation}
with the mass order $m_{\tilde{q}_1} \le \dots \le m_{\tilde{q}_6}$ for $q=u,d$, respectively. The rotation matrices appear in the couplings of squarks with other particles, and, in consequence, the flavour-violating elements will influence observables like decay widths or production and annihilation cross sections. Analytical expressions for couplings including squark generation mixing can, e.g., be found in Refs.\ \cite{NMFV_mSUGRA, NMFV_GMSB, NMFV_Squark1}. We shall discuss the relevant couplings for our analysis in more detail in Sec.\ \ref{sec3}.

A large variety of experimental measurements puts constraints on the parameter space of new physics models. Below are summarized all the constraints that will be considered (at the 95\% confidence level) in this study. The most important one for study this is naturally the relic density of cold dark matter. Combining data from the WMAP satellite and other cosmological measurements, the relic density of dark matter in the universe is constrained to \cite{WMAP}
\begin{equation}
	\Omega_{\rm CDM}h^2 = 0.1126 \pm 0.0036,
\label{eq:WMAP}
\end{equation}
where $h$ denotes the present Hubble expansion rate $H_0$ in units of 100 km s$^{-1}$\,Mpc$^{-1}$.

Then, searches for superpartners at LEP and Tevatron lead to the following mass limits for Higgs bosons, neutralinos, charginos, squarks, and gluinos: $m_{h^0} > 114.4$ GeV, $m_{\tilde{\chi}^0_1} > 46$ GeV, $m_{\tilde{\chi}^{\pm}_1} > 94$ GeV, $m_{\tilde{t}_1} > 96$ GeV, $m_{\tilde{g}} > 308$ GeV \cite{PDG}. Moreover, recent results from the Large Hadron Collider (LHC) lead to more stringent limits for squarks and gluinos within the constrained MSSM \cite{LHClimits}. These limits are based on the hypothesis of minimal flavour violation, so that we do not take them into account explicitly in the present study. Note, however, that the scenarios considered in the following feature rather heavy gluinos. For the lightest Higgs boson, we require $m_{h^0} > 111.4$~GeV, taking into account a theoretical uncertainty of 3 GeV \cite{SPheno}.

Moreover, precision measurements in the sector of D-, B-, and K-mesons constrain some of the flavour-violating elements in the mass matrices. In particular, flavour mixing involving the first generation of squarks is severely limited \cite{Gabbiani, Hagelin, Ciuchini}. We therefore focus on flavour mixing between the second and third generation squarks. 

The most relevant constraints on such flavour mixing are listed in Tab.\ \ref{tab1} together with the current experimental measurements and the theoretical error estimate. If not indicated otherwise, they are taken from Refs.\ \cite{PDG, HFAG}. They include branching ratios of rare decays, B-meson oscillation measurements, the electroweak $\rho$-parameter, and the anomalous magnetic moment of the muon. For the latter, taking into account recent calculations which bring the standard model theoretical expectation closer to the experimental measured value \cite{Bodenstein}, we use only the upper bound given in \cite{PDG} as a constraint. In the following study, the most important limits are imposed through the precise measurements of the rare decay $b\to s\gamma$ and the B-meson oscillation parameter $\Delta M_{B_s}$. 

\begin{table}
\caption{Experimental constraints on the MSSM parameter space, in particular on quark flavour violating elements.}
\bigskip
\begin{tabular}{|c|ccc|}
	\hline
	          & Exp.\ value & Exp.\ error & Theor.\ uncertainty \\
	\hline
	$10^4 \times {\rm BR}(b\to s\gamma)$ & $3.55$ & $\pm 0.26$ & $\pm 0.23$ \cite{bsgNLO}\\
    	$10^8 \times {\rm BR}(B_s\to \mu^+ \mu^-)$ & $<5.6$ \cite{LHCb}& & \\
	$\Delta M_{B_s}$ [ps$^{-1}$] & $17.77$ & $\pm 0.12$ & $\pm 3.3$ \cite{Deltambs} \\
	$\Delta \rho$ &$<0.0012$ \cite{Deltarho}& & \\
	$10^{11} \times \Delta a_{\mu}$ &$255$& $\pm 80$ & \\
	\hline
\end{tabular}
\label{tab1}
\end{table}

\section{Impact on the relic density of dark matter \label{sec3}}

The relic abundance in our universe of a stable dark matter candidate can be evaluated by solving the Boltzmann equation
\begin{equation}
	\frac{{\rm d}n}{{\rm d}t} ~=~ -3 H n - \langle \sigma_{\rm ann}v \rangle \left( n^2 - n^2_{\rm eq} \right) ,
\end{equation}	
where $n$ is the number density of the relic particle, $H$ the (time-dependent) Hubble expansion rate, and $n_{\rm eq}$ the number density in thermal equilibrium. All information concerning the particle physics model parameters is contained in the annihilation cross section $\sigma_{\rm ann}$ multiplied with the relative velocity $v$ of the annihilating particles. This product has to be convolved with the velocity distribution of the non-relativistic dark matter particle in order to obtain the thermally averaged cross section $\langle \sigma_{\rm ann}v \rangle$. 

Denoting the mass of the dark matter candidate by $m_0$ and taking into account a set of $N$ potentially co-annihilating particles with masses $m_i$ ($i=1,\dots,N$) such that $m_0 \leq m_1 \leq \dots \leq m_N$, the thermally averaged annihilation cross section can be written as \cite{GriestSeckel, EdsjoGondolo}
\begin{equation}
	\langle \sigma_{\rm ann}v \rangle ~=~ 
		\sum_{i,j=0}^N \langle\sigma_{ij}v_{ij}\rangle \frac{n^{\rm eq}_i n^{\rm eq}_j}{n_{\rm eq}^2} ~=~ 
	  	\sum_{i,j=0}^N \langle\sigma_{ij}v_{ij}\rangle \frac{g_i g_j}{g^2_{\rm eff}} 
		\left( \frac{m_i m_j}{m_0^2} \right)^{3/2} {\rm exp}\left\{-\frac{(m_i+m_j-2m_0)}{T}\right\}.
\label{eq:CoAnn}
\end{equation}
Here, $n^{\rm eq}_i$ denotes the equilibrium density of the particle $i$ and $i=0$ refers to the dark matter candidate. The cross sections $\sigma_{ij}$ relate to the different coannihilation processes within the ensemble of particles and $v_{ij}$ is the relative velocity between the particles $i$ and $j$. Moreover, $g_i$ denotes the number of degrees of freedom of particle $i$ and $g_{\rm eff}$ is a normalization factor.  

From Eq.\ (\ref{eq:CoAnn}) it becomes immediately clear that the mass differences between the annihilating particles play a crucial role. Due to the exponential suppression, the coannihilation between two given particles $i$ and $j$ will only lead to a significant contribution, if the two masses $m_i$ and $m_j$ are nearly degenerate \cite{GriestSeckel}. 

\begin{figure}
   \begin{center}
	\includegraphics[width=0.3\textwidth]{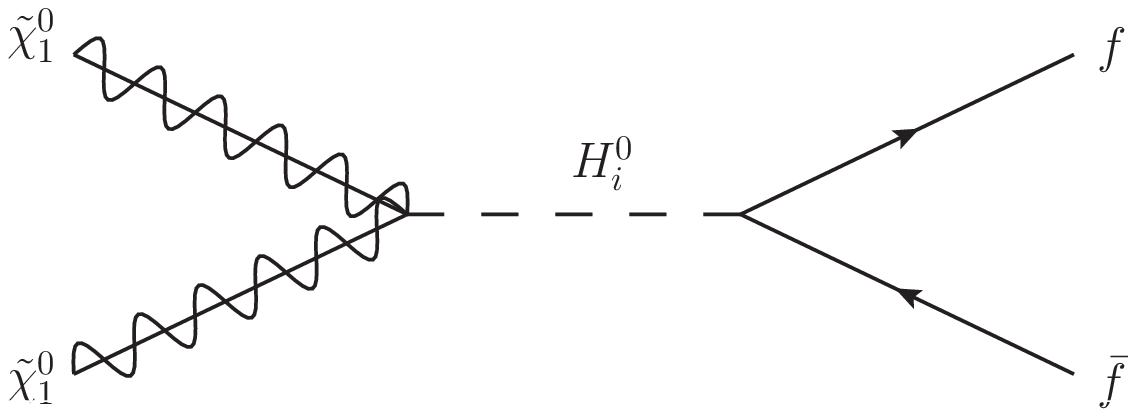}
	\includegraphics[width=0.3\textwidth]{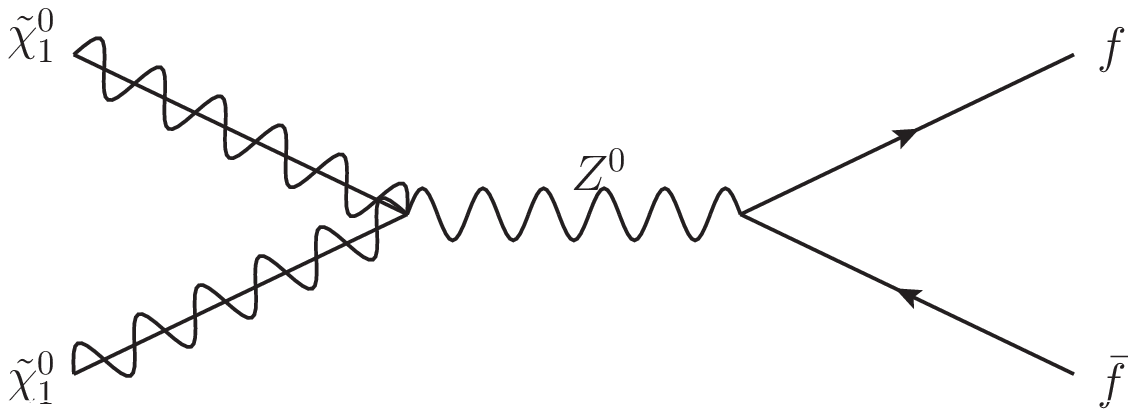}
	\includegraphics[width=0.3\textwidth]{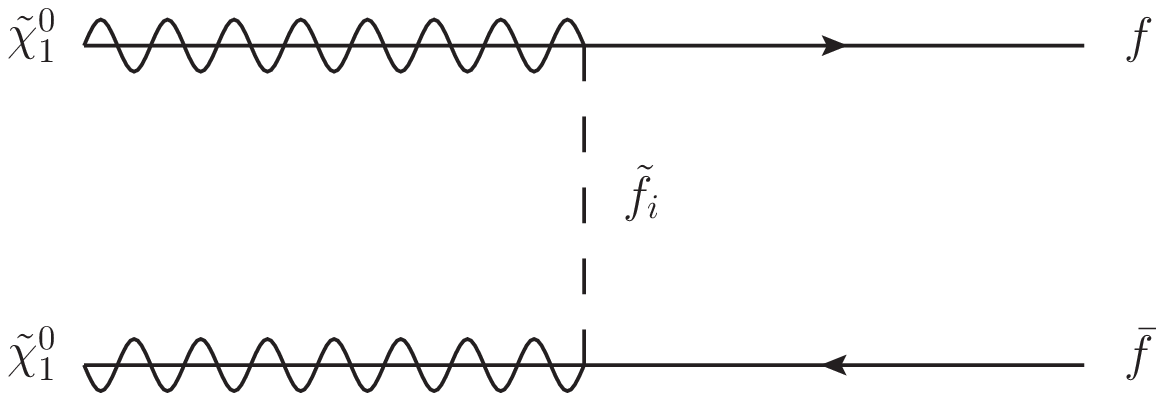}
   \end{center}
   \caption{Feynman diagrams for the annihilation of neutralinos into fermion pairs through the exchange of a neutral Higgs boson $H^0_i = h^0, H^0, A^0$ (left), a $Z^0$-boson (centre), or a sfermion (right). The corresponding $u$-channel diagram, obtained through crossing, is not shown.}
   \label{fig:diagrams1}
\end{figure}

In the following discussion, we assume that the lightest neutralino is the lightest supersymmetric particle (LSP) and therefore the dark matter candidate. In wide regions of the MSSM parameter space, the pair annihilation of two neutralinos into standard model particles is the dominant process. The diagrams for annihilation into quarks, i.e.\ where flavour violation in the (s)quark sector can become relevant, are shown in Fig.\ \ref{fig:diagrams1}. At the tree-level, squarks can then appear only in internal propagators in case of annihilation into quark-antiquark pairs, i.e.\ $\tilde{\chi}^0_1 \tilde{\chi}^0_1 \to q\bar{q}$ through the exchange of a squark in the $t$- or $u$-channel \cite{DM_mSUGRA, DM_NUHM}. 

Going beyond minimal flavour violation, the mass splitting of the involved squarks is increased due to the additional off-diagonal elements in the mass matrix. In particular, the lightest squark mass eigenstate (purely stop-like in the CMSSM with MFV) becomes lighter with increasing flavour mixing. Its contributions to neutralino pair annihilation through $t$- or $u$-channel exchange are therefore enhanced.

Apart from the impact on the squark mass eigenvalues, the flavour-violating terms discussed in Sec.\ \ref{sec2} directly affect the neutralino-squark-quark coupling, which is present in the $t$- or $u$-channel diagram. The analytical expressions for the left- and right-handed parts of this coupling are given by \cite{NMFV_mSUGRA}
\begin{eqnarray}
	\label{eq:NeuSqQ}
	L_{\tilde{\chi}^0_i \tilde{u}_j u_k} &=& \Big[ \left( e_u - T_u \right) \sin\theta_W {\cal N}_{i1} + T_u \cos\theta_W {\cal N}_{i2} \Big] {\cal R}^{\tilde{u}*}_{jk} + \frac{m_{u_k} \cos\theta_W}{2m_W \cos\beta} {\cal N}_{i4} {\cal R}^{\tilde{u}*}_{j(k+3)}, \\
	L_{\tilde{\chi}^0_i \tilde{d}_j d_k} &=& \Big[ \left( e_d - T_d \right) \sin\theta_W {\cal N}_{i1} + T_d \cos\theta_W {\cal N}_{i2} \Big] {\cal R}^{\tilde{d}*}_{jk} + \frac{m_{d_k} \cos\theta_W}{2m_W \sin\beta} {\cal N}_{i3} {\cal R}^{\tilde{d}*}_{j(k+3)}, \\
	-R^*_{\tilde{\chi}^0_i \tilde{u}_j u_k} &=& e_u \sin\theta_W {\cal N}_{i1} {\cal R}^{\tilde{u}}_{jk} + \frac{m_{u_k} \cos\theta_W}{2m_W \cos\beta} {\cal N}_{i4} {\cal R}^{\tilde{u}}_{j(k+3)}, \\
	-R^*_{\tilde{\chi}^0_i \tilde{d}_j d_k} &=& e_d \sin\theta_W {\cal N}_{i1} {\cal R}^{\tilde{d}}_{jk} + \frac{m_{d_k} \cos\theta_W}{2m_W \sin\beta} {\cal N}_{i3} {\cal R}^{\tilde{d}}_{j(k+3)}, 
\end{eqnarray}
with the same notations as in Sec.\ \ref{sec2}. Flavour mixing effects arise through the squark rotation matrix ${\cal R}^{\tilde{q}}$ ($q=u,d$). This can allow for new annihilation channels, that are closed in the case of minimal flavour violation. Such channels can, e.g., be $\tilde{\chi}\tilde{\chi} \to c\bar{c}$ through exchange of a squark $\tilde{u}_1$ which is now a mixture of $\tilde{c}$ and $\tilde{t}$. In the case of MFV, this final state is only possible through exchange of a heavier $\tilde{c}$ and therefore suppressed. Another example is annihilation into a mixed final state, $\tilde{\chi}\tilde{\chi} \to c\bar{t}$, which is forbidden in MFV. The discussed enhancements and new channels increase the total annihilation cross section, which in turn decreases the predicted relic density of the neutralino. The diagrams with $s$-channel exchange of a Higgs or gauge boson remain insensitive to squark flavour mixing.

\begin{figure}
\begin{center}
\includegraphics[width=0.3\textwidth]{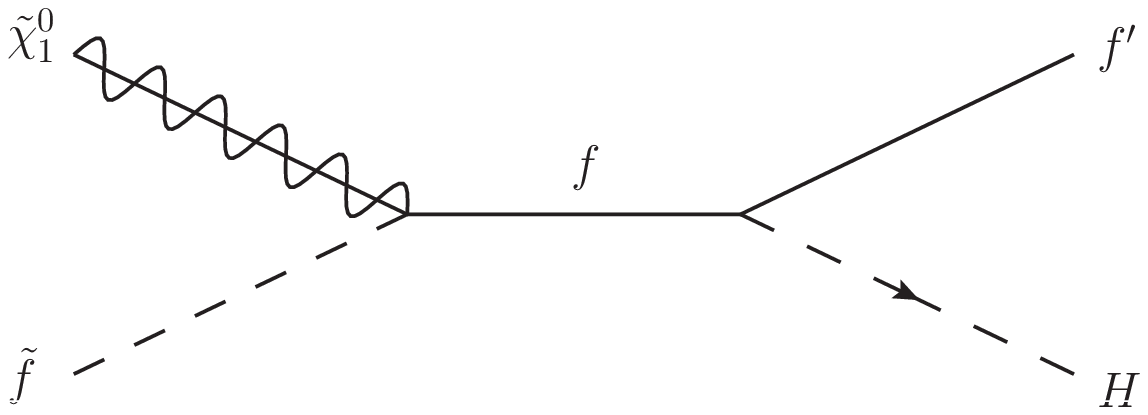}
\includegraphics[width=0.3\textwidth]{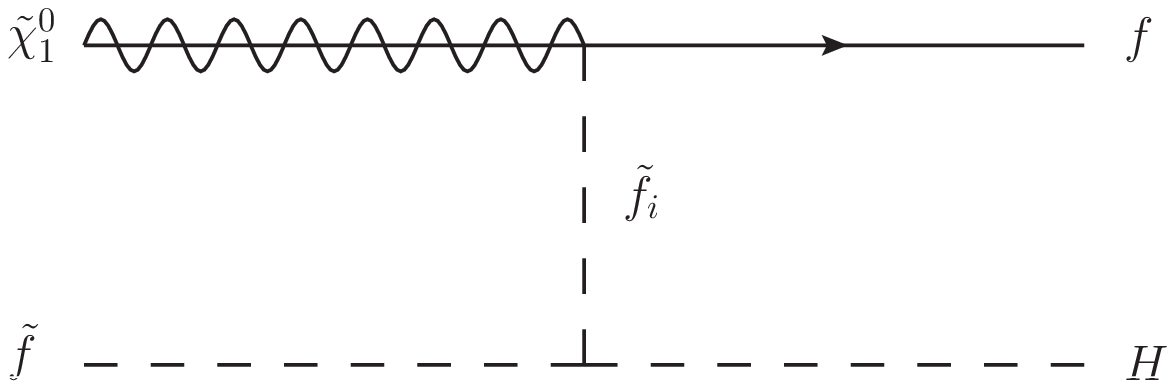}
\includegraphics[width=0.3\textwidth]{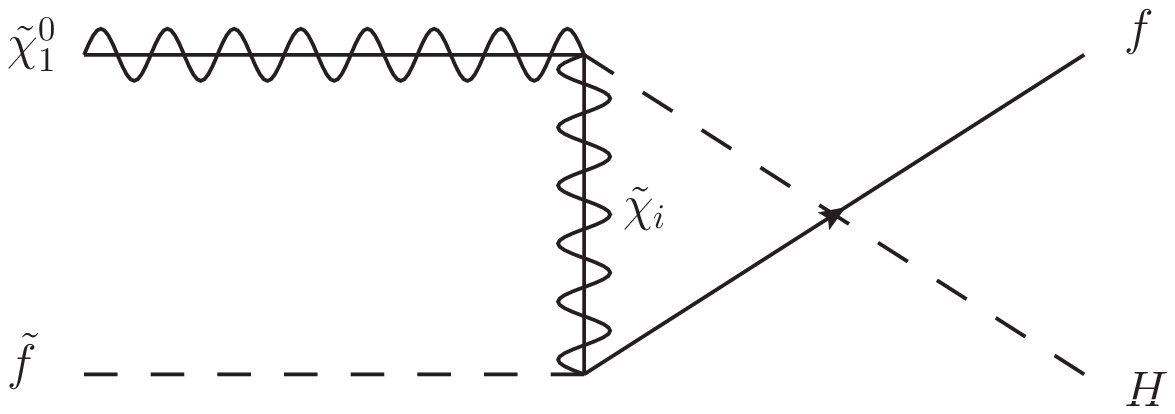}
\bigskip
\\
\includegraphics[width=0.3\textwidth]{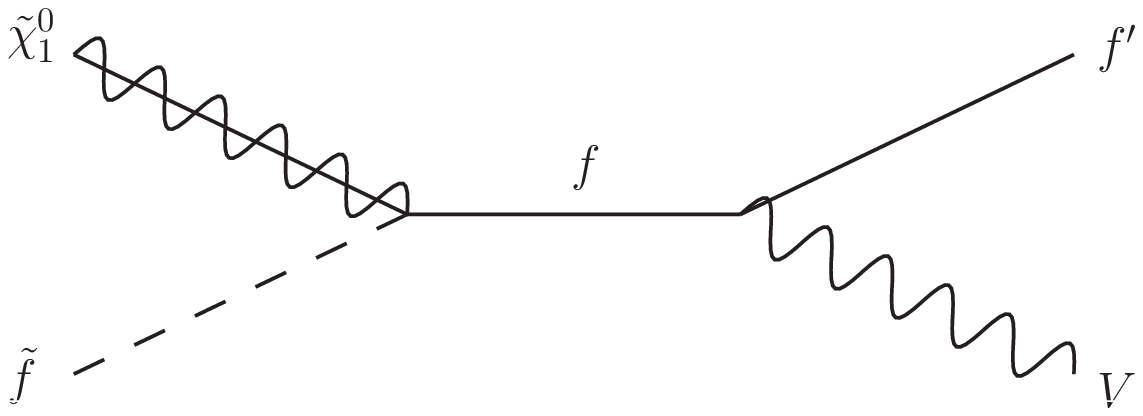}
\includegraphics[width=0.3\textwidth]{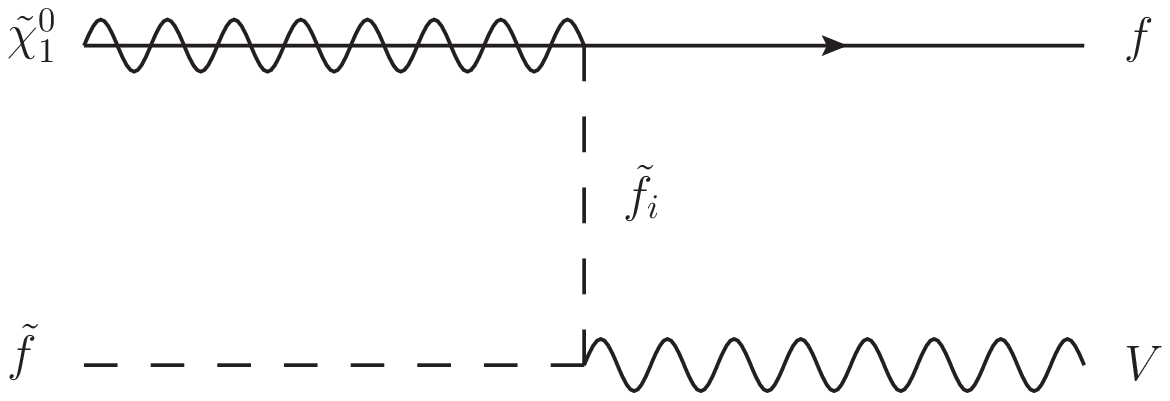}
\includegraphics[width=0.3\textwidth]{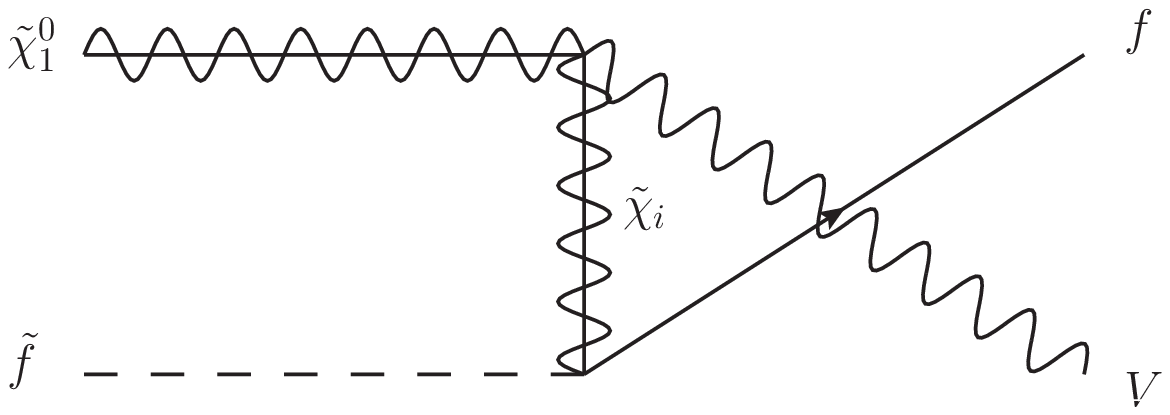}
\end{center}
\caption{Feynman diagrams for the coannihilation of a neutralino with a sfermion into a fermion together with a Higgs boson $H=h^0,H^0,A^0,H^{\pm}$ (top) or a gauge boson $V=\gamma, Z^0, W^{\pm}, g$ (bottom). These processes proceed through the exchange of a fermion (left), a sfermion (centre) or a gaugino (right). The $u$-channel diagram is not present in case of the gluonic final state.}
\label{fig:diagrams2}
\end{figure}

Let us now turn to the case of neutralino-squark coannihilation. The possible final states are a quark together with a Higgs or a gauge boson. The relevant Feynman diagrams at the tree-level are depicted in Fig.\ \ref{fig:diagrams2}. The main impact from non-minimal flavour violation will be through the modified squark mass spectrum. As already stated above, the mass difference between neutralino and the lightest squark enters the calculation of the corresponding coannihilation cross section exponentially. When the squark mass approaches the neutralino mass due to increasing flavour mixing, this can significantly enhance the contribution from the corresponding coannihilation with respect to the case of minimal flavour violation. 

Again, also the flavour-violating couplings can have subdominant effects on the coannihilation processes. Each of the diagrams depicted in Fig.\ \ref{fig:diagrams2} contains the squark-quark-neutralino coupling already discussed above. Moreover, the couplings of squarks to Higgs- and massive gauge bosons are sensitive to flavour-violating effects. In the mass eigenstate basis, the couplings of squarks to a $Z^0$-boson are given by \cite{NMFV_Squark1}
\begin{eqnarray}
  C_{Z^0 \tilde{q}_j \tilde{q}_k} &=&
    - i \frac{g_2}{\cos \theta_W} (p_j + p_k)_{\mu} \left[ \sum_{i=1}^3 I_q {\cal R}^{\tilde{q}*}_{ij} {\cal R}^{\tilde{q}}_{ik}
      - e_{q} \sin^2\theta_W \delta_{jk}
      \right] 
\label{eq:SqSqZ}
\end{eqnarray}
for $q=u,d$. Here, $p_j$ and $p_k$ denote the momentum of $\tilde{q}_j$ and $\tilde{q}_k$, respectively. The interactions of squarks with a photon or a gluon are flavour-diagonal and are therefore not discussed in detail here.

The couplings of two up-type squarks with the light scalar Higgs boson are given by \cite{NMFV_Squark1}
\begin{eqnarray}
	 C_{h^0 \tilde{u}_j \tilde{u}_k} & = & -\frac{g_2}{2 m_W}\ \sum_{i=1}^3
	    \bigg[ m_W^2 \sin(\alpha+\beta) \Big[
	      (1-\frac{1}{3} \tan^2\theta_W) {\cal R}^{\tilde{u}}_{ji} {\cal R}^{\tilde{u}*}_{ki}
	      + \frac{4}{3} \tan^2\theta_W {\cal R}^{\tilde{u}}_{j(i+3)} {\cal R}^{\tilde{u}*}_{k(i+3)}
	      \Big] \nonumber \\[0.2cm]
	    & & + 2 \frac{\cos\alpha}{\sin\beta} \Big[
	      {\cal R}^{\tilde{u}}_{ji}\ m^2_{u_i} {\cal R}^{\tilde{u}*}_{ki}
	      + {\cal R}^{\tilde{u}}_{j(i+3)} m^2_{u_i} {\cal R}^{\tilde{u}*}_{k(i+3)}
	      \Big] + \frac{\sin\alpha}{\sin\beta} \Big[
	      \mu^* {\cal R}^{\tilde{u}}_{j(i+3)} m_{u_i} {\cal R}^{\tilde{u}*}_{ki}
	      + \mu {\cal R}^{\tilde{u}}_{ji} m_{u_i} {\cal R}^{\tilde{u}*}_{k(i+3)}
	      \Big] \nonumber \\[0.2cm]
	    & & + \frac{\cos\alpha}{\sin\beta}\, \frac{v_u}{\sqrt2} \sum_{l=1}^3 \Big[
	      {\cal R}^{\tilde{u}}_{j(i+3)}\ ({T}_U)_{il}\ {\cal R}^{\tilde{u}*}_{kl}
	      + {\cal R}^{\tilde{u}}_{ji}\ ({T}_U^\dagger)_{il}\ {\cal R}^{\tilde{u}*}_{k(l+3)}
	    \Big] \bigg] .
\label{eq:SqSqH}
\end{eqnarray}
From this expression, the coupling to the heavy scalar Higgs is obtained through the replacements $h^0 \to H^0$ and $\alpha \to \alpha+\pi/2$. Moreover, couplings of down-type squarks to the neutral scalar Higgses are obtained by replacing $\tilde{u}_i \to \tilde{d}_i$ and $\sin\beta \to \cos\beta$. Finally, the couplings of up-type squarks to a pseudoscalar Higgs-boson are given by \cite{NMFV_Squark1}
\begin{equation}
	 C_{A^0 \tilde{u}_j \tilde{u}_k} = -i \frac{g_2}{2 m_W}\ \sum_{i=1}^3 \Big[
	        \mu^* {\cal R}^{\tilde{u}}_{j(i+3)} m_{u_i} {\cal R}^{\tilde{u}*}_{ki} + \cot\beta \frac{v_u}{\sqrt2}
	        \sum_{l=1}^3 {\cal R}^{\tilde{u}}_{j(i+3)} (T_U)_{il} {\cal R}^{\tilde{u}*}_{kl} + \textnormal{h.c} \Big].
\label{eq:SqSqA}
\end{equation}
Again, the expressions for down-type squarks can easily be obtained through $\tilde{u}_i \to \tilde{d}_i$ and $\cot\beta \to \tan\beta$.

The effects of the modified mass eigenvalues and the modified couplings are superimposed. Since the two effects are linked together through their same origin (see Eq.\ (\ref{eq:massdiag})), the separate impacts on the (co)annihilation cross section and the neutralino relic density cannot be disentangled. However, some general features can be expected. The effect of the modified squark mass eigenvalues on coannihilation is expected to be stronger than in the case of neutralino pair annihilation due to the exponential factor already mentioned above. Moreover, the squark is here an external particle, and the impact of its mass on the phase space is more important than the mass in the $t$- or $u$-channel propagator. 

The impact of the modified flavour contents of the involved squarks, i.e.\ the effect of the rotation matrix in the coupling, is expected to be smaller than the mass effect. This is again due to the exponential factor in Eq.\ (\ref{eq:CoAnn}). Note also that, the mixing being unitary, the newly opened channels can be (partially) compensated by the simultaneous diminution of other contributions. The compensating contribution can, however, turn out to be forbidden in specific kinematical configurations and the impact of the new contributions can be significant. This is in particular the case when the neutralino is too light to annihilate into top-quark pairs, i.e.\ for $m_{\tilde{\chi}_1^0} < m_t$. The flavour violating elements lead then to a $\tilde{c}$ admixture in the lightest squark, which then allows for neutralino pair annihilation into top and charm quarks.

Note that there can also be coannihilation of a neutralino with an up-(down-)type squark into a charged Higgs boson $H^{\pm}$ or a W-boson together with a down-(up-)type quark. In this case, the $u$-channel diagram includes a chargino propagator and in consequence the corresponding chargino-squark-quark coupling, while the $s$- and $t$-channel diagrams involve couplings of up- and down-type squarks to the charged Higgs or W-boson. Analytical expressions for these couplings can be found in Refs.\ \cite{NMFV_mSUGRA, NMFV_Squark1}. Since they are rather similar (with obvious replacements, e.g., concerning gaugino mixing) to the interactions given in Eqs.\ (\ref{eq:NeuSqQ}) to (\ref{eq:SqSqH}), they are not displayed in detail here. Note, however, that these couplings explicitly depend on the CKM-matrix. The general argumentation given above remains unchanged.

\section{Numerical analysis \label{sec4}}

The following numerical analyses are mainly based on the constrained MSSM with the five parameters $m_0$, $m_{1/2}$, $A_0$, $\tan\beta$, and sgn($\mu$). We also consider variants of this model featuring non-universal Higgs or gaugino masses. Starting from the high-scale parameters, the soft-breaking terms at the scale $Q=1$~TeV \cite{SPA} are obtained through renormalization group running using the public program {\tt SPheno~3} \cite{SPheno}. At the same scale, we introduce the non-diagonal entries in the squark mass matrices as discussed in Sec.\ \ref{sec2}. The physical mass spectrum is then calculated again using {\tt SPheno}, which takes into account the general flavour structure. The same code is also used for the evaluation of the constraining observables mentioned in Sec.\ \ref{sec2}, again taking into account squark generation mixing. For the standard model parameters, we refer the reader to Ref.\ \cite{PDG}. The pole mass of the top-quark is taken to be $m_{\rm top} = 173.1$~GeV according to recent measurements from D0 and CDF \cite{TopMass}. The CKM-matrix is taken in the usual Wolfenstein parametrization with the recent values $\lambda=0.2253$, $A=0.808$, $\bar{\rho}=0.132$, and $\bar{\eta}=0.341$
\cite{PDG}.

Making use of the SUSY Les Houches Accord \cite{SLHA}, the mass spectrum and related mixing parameters are transferred to the public program {\tt micrOMEGAs~2.4} \cite{micrOMEGAs} in order to evaluate the relic density of the neutralino. The calculation of the annihilation cross section is done by the program {\tt CalcHEP} \cite{CalcHEP}, where we have implemented the MSSM with squark generation mixing as discussed in Sec.\ \ref{sec2}. The corresponding model files have been obtained using the package {\tt SARAH} \cite{SARAH}. We also include important effects from the running strong coupling constant and running quark masses, as they are also included in the default implementation of the MSSM in {\tt micrOMEGAs} / {\tt CalcHEP}.

\subsection{Constrained MSSM \label{sec4a}}

\begin{figure}
   \begin{center}
	\includegraphics[width=0.45\textwidth]{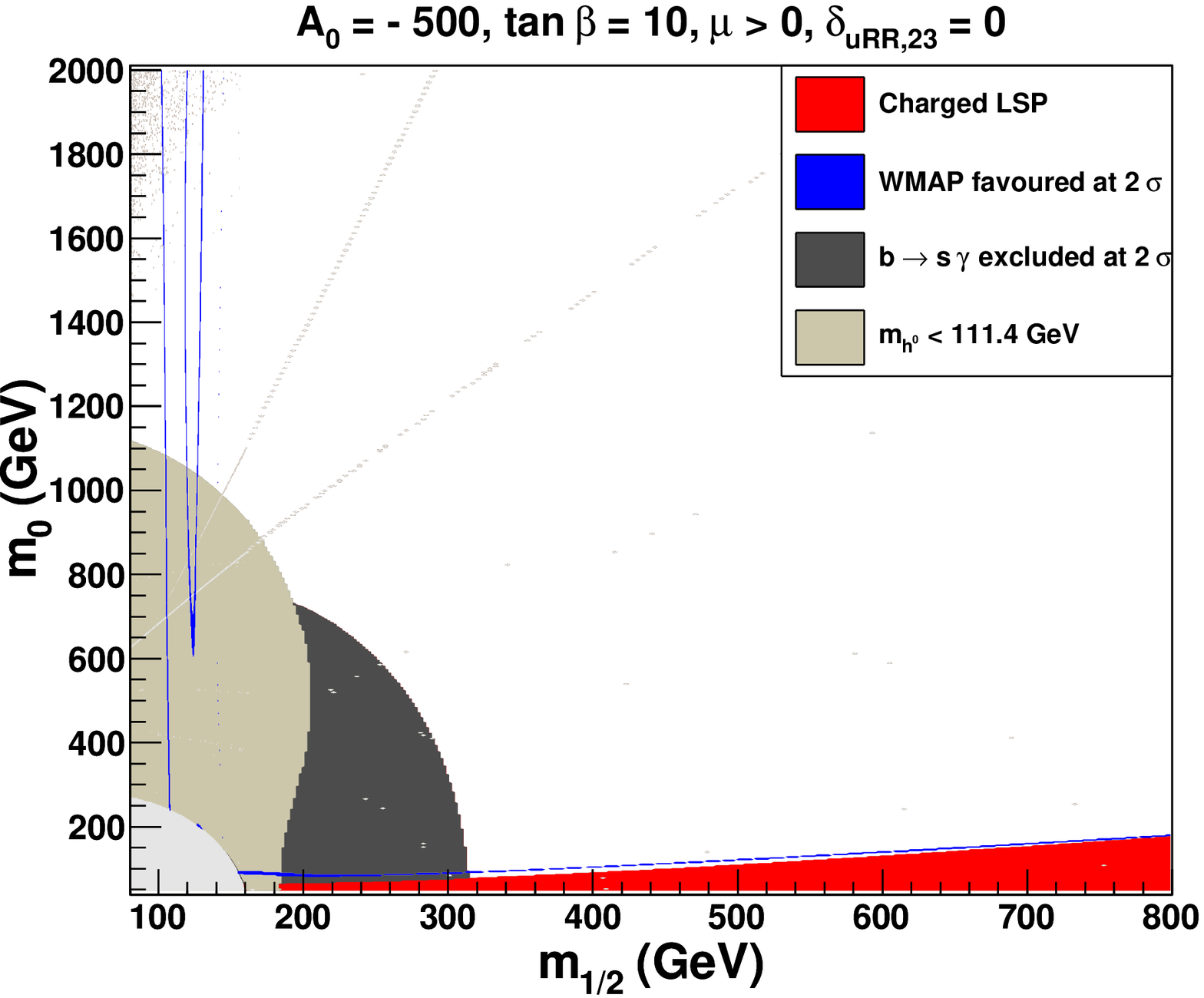}
	\includegraphics[width=0.45\textwidth]{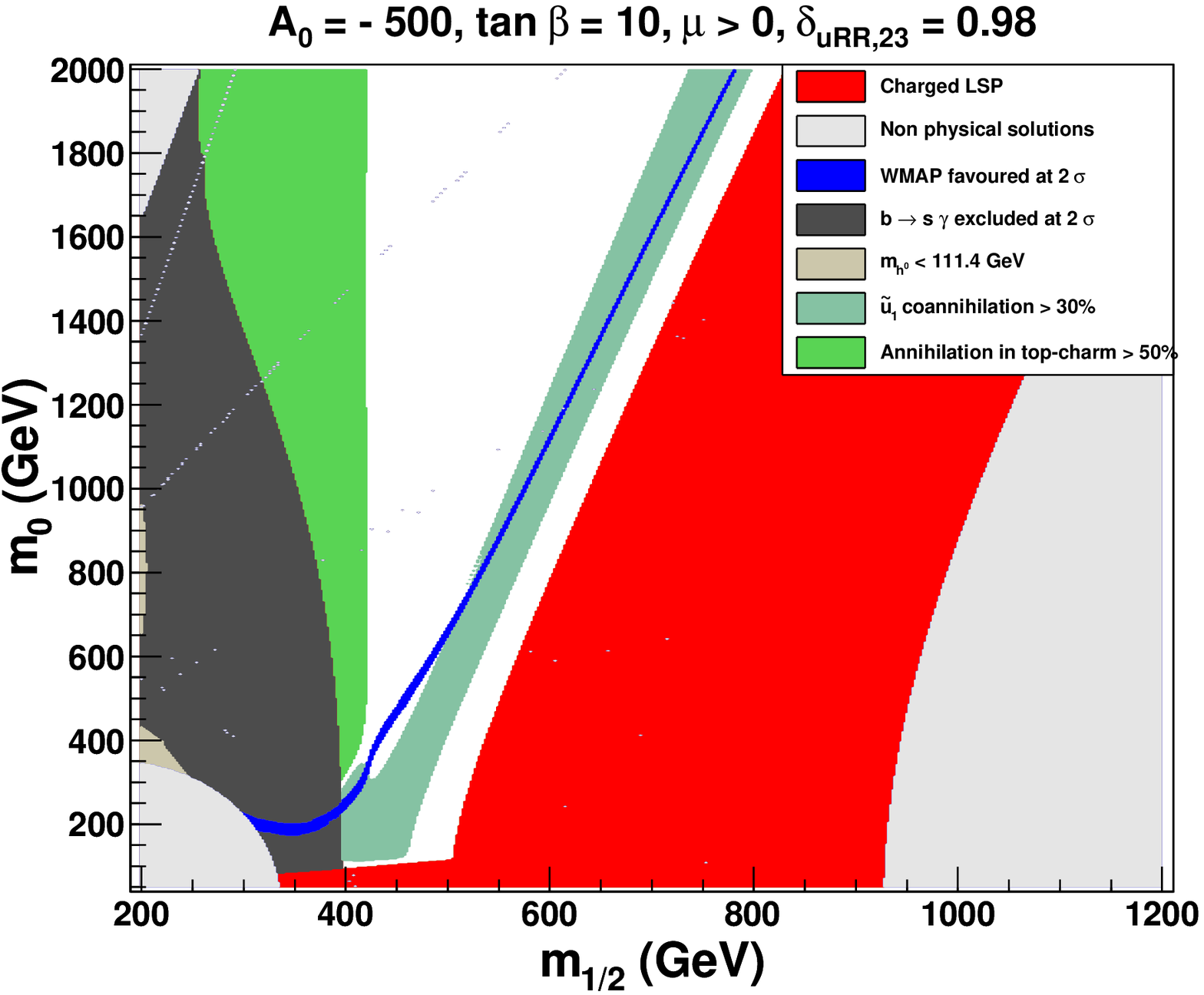}
   \end{center}
   \caption{Cosmologically favoured region and related exclusion limits in the ($m_0$, $m_{1/2}$) plane of the CMSSM for $\delta^{u,\rm RR}_{23} = 0$ (left) and $\delta^{u,\rm RR}_{23} = 0.98$ (right).}
   \label{fig:CMSSM1}
\end{figure}

In order to illustrate the numerical influence of flavour violating elements, we start by analyzing the neutralino relic density within the constrained MSSM (CMSSM), where we allow for flavour violation between the second and third generation of up-type squarks in the right-right chiral sector. In Fig.\ \ref{fig:CMSSM1}, we show typical scans of the $m_0$-$m_{1/2}$ plane for fixed values of $A_0 = -500$ GeV and $\tan\beta=10$ and for positive values of $\mu$. The cosmologically favoured region of parameter space according to Eq.\ (\ref{eq:WMAP}) together with the relevant constraints discussed in Sec.\ \ref{sec2} are shown for the case of minimal flavour violation (MFV, $\delta^{u,\rm RR}_{23} = 0$) and for the case of important off-diagonal elements, $\delta^{u,\rm RR}_{23} = 0.98$.

In the case of MFV, the most stringent constraints on this parameter plane are due to a charged dark matter candidate (low $m_0$), tachyonic solutions of the renormalization group equations (high $m_0$ and low $m_{1/2}$) as well as the constraints from $b\to s\gamma$ and the lightest Higgs mass (low mass region). The cosmologically favoured region of parameter space is divided into several distinct regions: the so-called focus point region (high $m_0$, not visible here), the resonance of the light Higgs boson (low $m_{1/2}$ and moderate $m_0$), and the coannihilation region (close to the exclusion due to a charged dark matter candidate), where the neutralino mass is close to the stau mass.

In the corresponding figure for the NMFV-case, we depict the same constraints together with the relative contribution from new (co)annihilation channels as discussed in Sec.\ \ref{sec3}. In this case, this involves neutralino pair annihilation into a mixed charm-top final state and coannihilation of a neutralino with the lightest squark $\tilde{u}_1$.

\begin{figure}
   \begin{center}
	\includegraphics[width=0.45\textwidth]{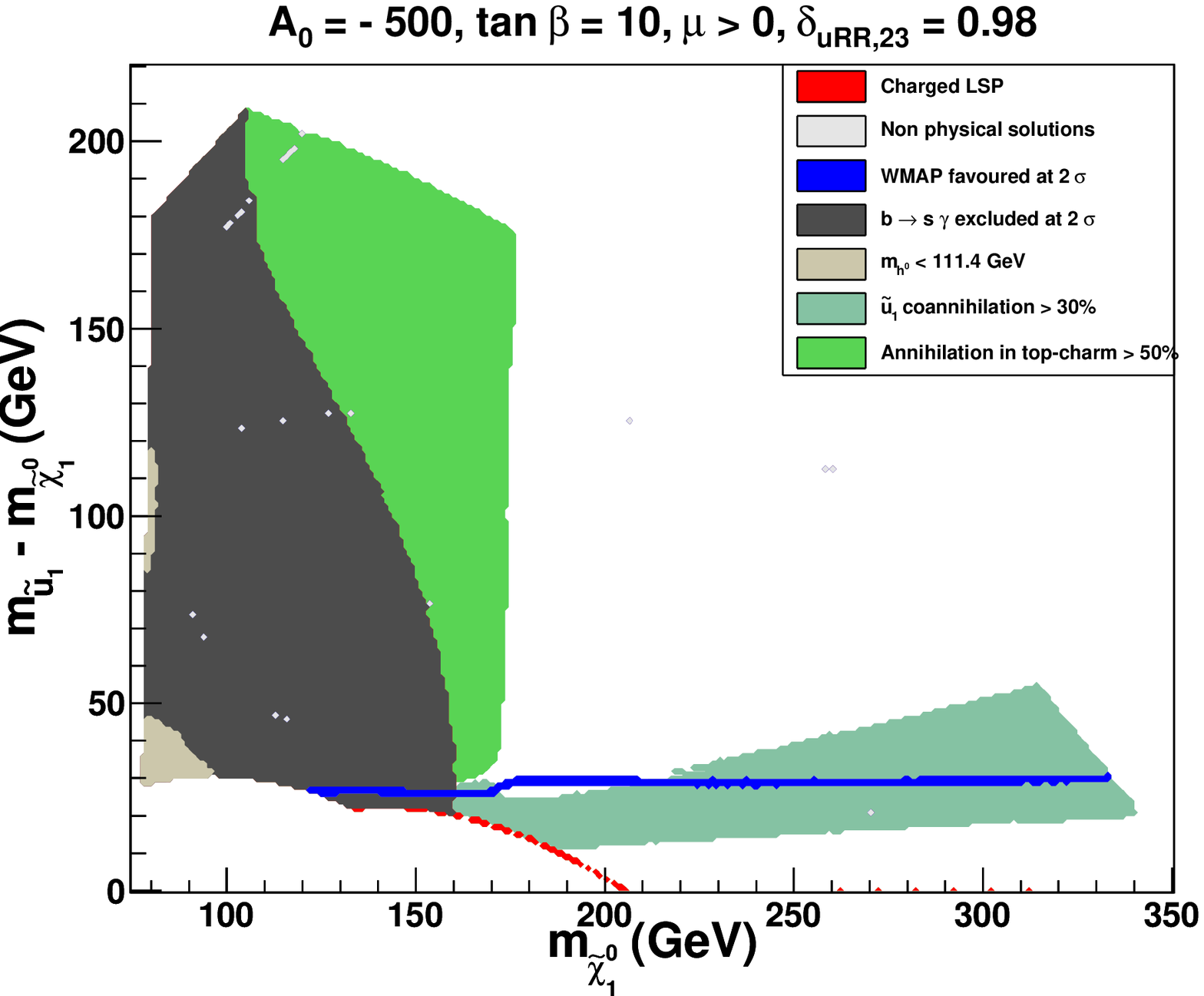}
	\includegraphics[width=0.45\textwidth]{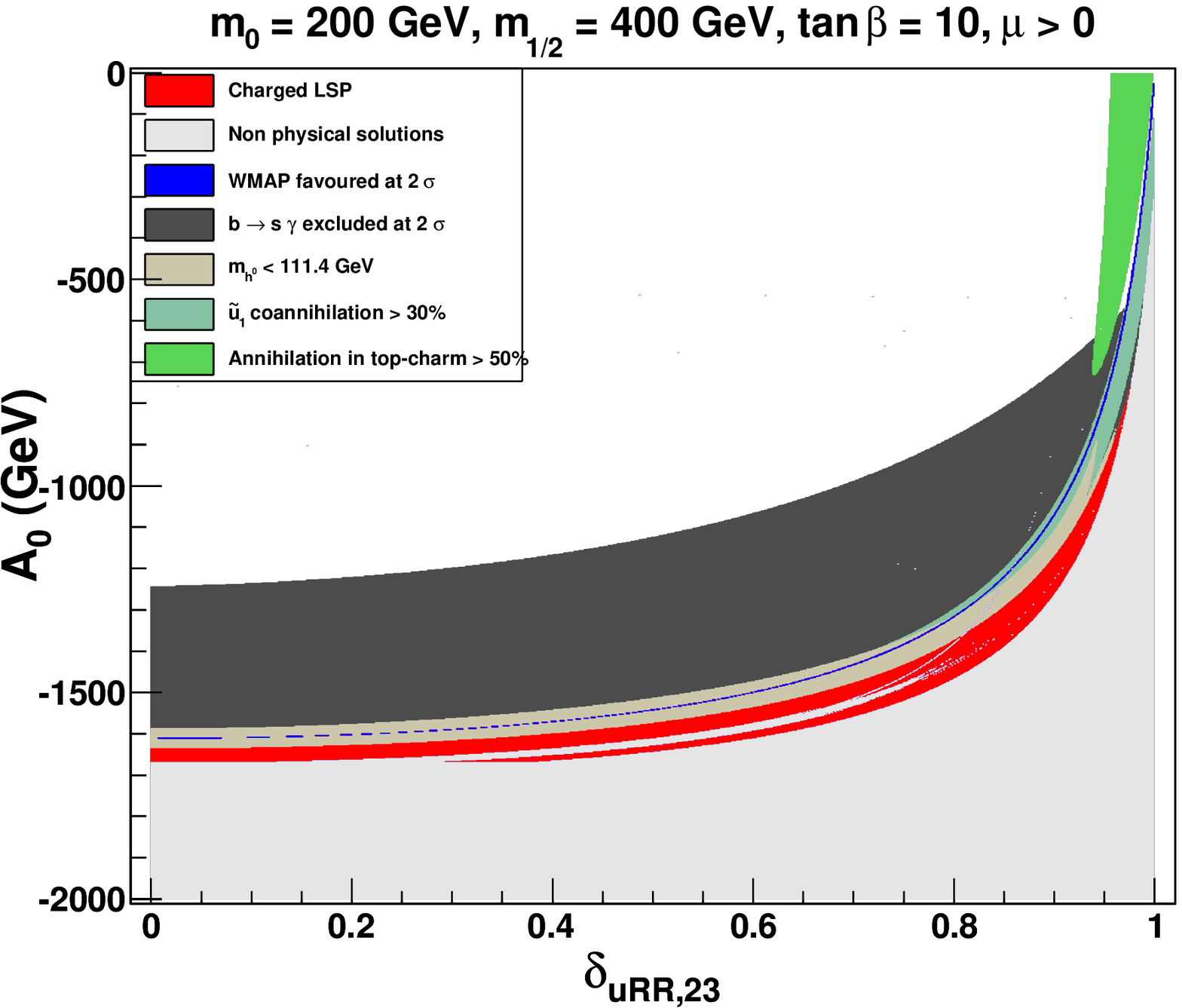}
   \end{center}
   \caption{Cosmologically favoured region and related exclusion limits for $\delta^{u,\rm RR}_{23} = 0.98$ in the ($m_{\tilde{\chi}^0_1}$, $m_{\tilde{u}_1}-m_{\tilde{\chi}^0_1}$) plane for fixed $A_0=-500$~GeV and $\tan\beta=10$ (left) and in the ($\delta^{u,\rm RR}_{23}$, $A_0$) plane for fixed $m_0=200$ GeV and $m_{1/2}=400$ GeV (right).}
   \label{fig:CMSSM2}
\end{figure}

In the latter corresponding region ($m_{1/2} \gtrsim 450$ GeV), where the relic density constraint is fulfilled, the mass difference between the lightest squark and the neutralino is about $30$ GeV, as can be seen from the left panel of Fig.\ \ref{fig:CMSSM2}, where we show the cosmologically favoured regions of parameter space in the plane of the physical masses. The dominant annihilation processes are then $\tilde{\chi}^0_1 \tilde{u}_1 \to g t$ ($30 \%$) and $\tilde{u}_1 \tilde{u}_1 \to g g$ ($25 \%$). Two other important processes are neutralino annihilation into pairs of top quarks ($10 \%$), and $\tilde{\chi}^0_1 \tilde{u}_1 \to g c$ ($15 \%$). Note that the presence of a charm quark in the final state is a genuine effect of flavour violation. Indeed, as a consequence of the off-diagonal elements in squark mass matrices, the lightest up-type squark is here a mixing of $\tilde{t}_R$ and $\tilde{c}_R$ (with a small admixture of $\tilde{t}_L$), opening up the (co)annihilation into charm-quarks.

For lower masses (e.g.\ $m_0 \sim 200$ GeV and $m_{1/2} \sim 400$ GeV), coannihilation processes such as $\tilde{\chi}^0_1 \tilde{u}_1 \to g t/c$ are still important ($20 \%$). However, the squark being much lighter ($m_{\tilde{u}_1} \sim 190$ GeV), the squark pair annihilation $\tilde{u}_1 \tilde{u}_1 \to gg$ is now subdominant. Moreover, the neutralino mass of $m_{\tilde{\chi}_1^0} \sim 160$ GeV (see Fig.\ \ref{fig:CMSSM2} left) forbids annihilation into top quark pairs. As a consequence, the flavour violating process $\tilde{\chi}^0_1 \tilde{\chi}^0_1 \to t \bar{c} (c \bar{t})$, which is kinematically allowed and enhanced by the rather light squark in the $t$-channel propagator, becomes important ($40 \%$). This is represented by the green area in the left part of the plot. Notice the cut at $m_{1/2} \approx 420$ GeV, which corresponds to $m_{\tilde{\chi}^0_1} \approx m_t$. For $m_{\tilde{\chi}^0_1} > m_t$, neutralino annihilation into top quark pairs is kinematically allowed, and the $t\bar{c} (c\bar{t})$ final state is suppressed. This can also be seen in relation to the physical neutralino and squark masses in Fig.\ \ref{fig:CMSSM2} left. For low $m_{1/2}$ but large $m_0$, the squark being heavier, coannihilation is not relevant and neutralino annihilation into $t\bar{c} (c\bar{t})$ is less important. Therefore, even if the relative contribution of this channel is still important, its absolute contribution is not large enough to satisfy the relic density constraint.

In the region excluded by $\textnormal{BR}(b \to s \gamma)$ most of the deviation from the standard model value comes from large negative chargino contributions due to the smallness of the stop and/or chargino mass. There is, however, no significant effect coming from the flavour violating parameter $\delta^{u,\rm RR}_{23}$, since $\textnormal{BR}(b \to s \gamma)$ constrains mainly flavour violation in the left-left sector.

Let us now discuss the interplay of helicity mixing and additional flavour mixing. The former is induced through the trilinear matrices $T_{U}$ (see Eq.\ (\ref{eq:RLu})) and thus the GUT-scale parameter $A_0$, while the latter is included at the electroweak scale through the parameter $\delta^{u,\rm RR}_{23}$. In the case of MFV, i.e.\ for $\delta^{u,\rm RR}_{23}=0$, a rather large $|A_0|$ is needed in order to decrease the stop mass close to the neutralino mass, and therefore allow for efficient coannihilation. For sizeable additional flavour mixing, the coannihilation is important already for lower values of $A_0$, since the squark mass splitting is then increased by the off-diagonal elements in the mass matrix. 

This is illustrated in the right graph of Fig.\ \ref{fig:CMSSM2}, where the constraints, cosmologically favoured regions, and different contributions to the annihilation cross section are shown in the ($A_0$,$\delta^{u,\rm RR}_{23}$) plane. The mass splitting of the squarks depends strongly on both of these parameters, which therefore have a competitive effect on the light stop mass. As a consequence, as explained above, one of these parameters has to be large in order to allow for an important coannihilation contribution. On the other hand, the flavour violating effects are only related to $\delta^{u,\rm RR}_{23}$. Therefore the flavour violating neutralino annihilation processes depend mainly on this parameter. The only possibility to satisfy simultaneously the relic density and BR$(b \to s \gamma)$ constraints is for very large $\delta^{u,\rm RR}_{23}$ and a rather low $A_0$. This is explained by the strong dependance of BR$(b \to s \gamma)$ on the squark mass spectrum, and therefore on $A_0$. Contrary, and as explained above, BR$(b \to s \gamma)$ does not depend on any flavour mixing among right up-type squarks, and the mass effects become important only for very large values of $\delta^{u,\rm RR}_{23}$. It has been checked that the other constraints described in Tab.\ \ref{tab1} are fulfilled in the whole parameter space shown in Figs.\ \ref{fig:CMSSM1} and \ref{fig:CMSSM2}. Moreover the calculated spectrum  is compatible with the mass limits given in Sec.\ \ref{sec2}, except for the stop in some regions where it is the LSP (i.e. already excluded).

\begin{figure}
   \begin{center}
	\includegraphics[width=0.45\textwidth]{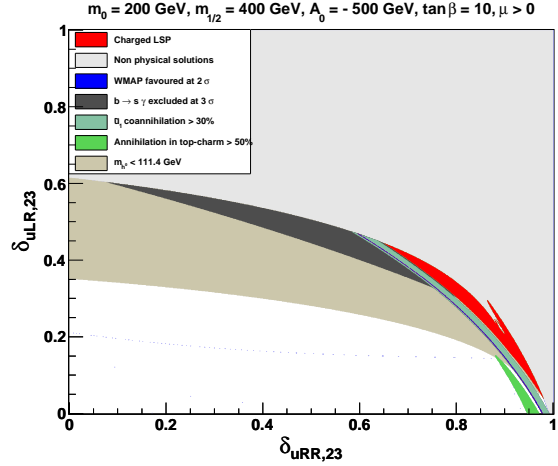}
   \end{center}
   \caption{Cosmologically favoured region and related exclusion limits in the $(\delta_{23}^{u,RR},\delta_{23}^{u,LR})$ plane for fixed $m_0=200$ GeV, $m_{1/2}=400$ GeV, and $A_0=-500$ GeV.}
   \label{fig:CMSSM2_2}
\end{figure}

Next, we study the possibility that not only the parameter $\delta_{23}^{u,RR}$ is large, {\em i.e.} of ${\cal O}(1)$, while all others are small, which might not be very natural. We therefore show in Fig.\ \ref{fig:CMSSM2_2} the cosmologically favoured region and related exclusion limits in the $(\delta_{23}^{u,RR},
\delta_{23}^{u,LR})$ plane for fixed $m_0=200$ GeV, $m_{1/2}=400$ GeV, and $A_0=-500$ GeV. We observe that the second flavour-violating parameter $\delta_{23}^{u,LR}$ can reach values up to $0.15$ before being constrained by the lower Higgs mass bound of $111.4$ GeV. Similarly, the RL and LL parameters (not shown) are restricted by the FCNC process $b\to s\gamma$ to values below $0.15$ and $0.1$, respectively, as would be the LR parameter if one applied this limit at the two (not three) sigma level.

\begin{figure}
   \begin{center}
	\includegraphics[width=0.45\textwidth]{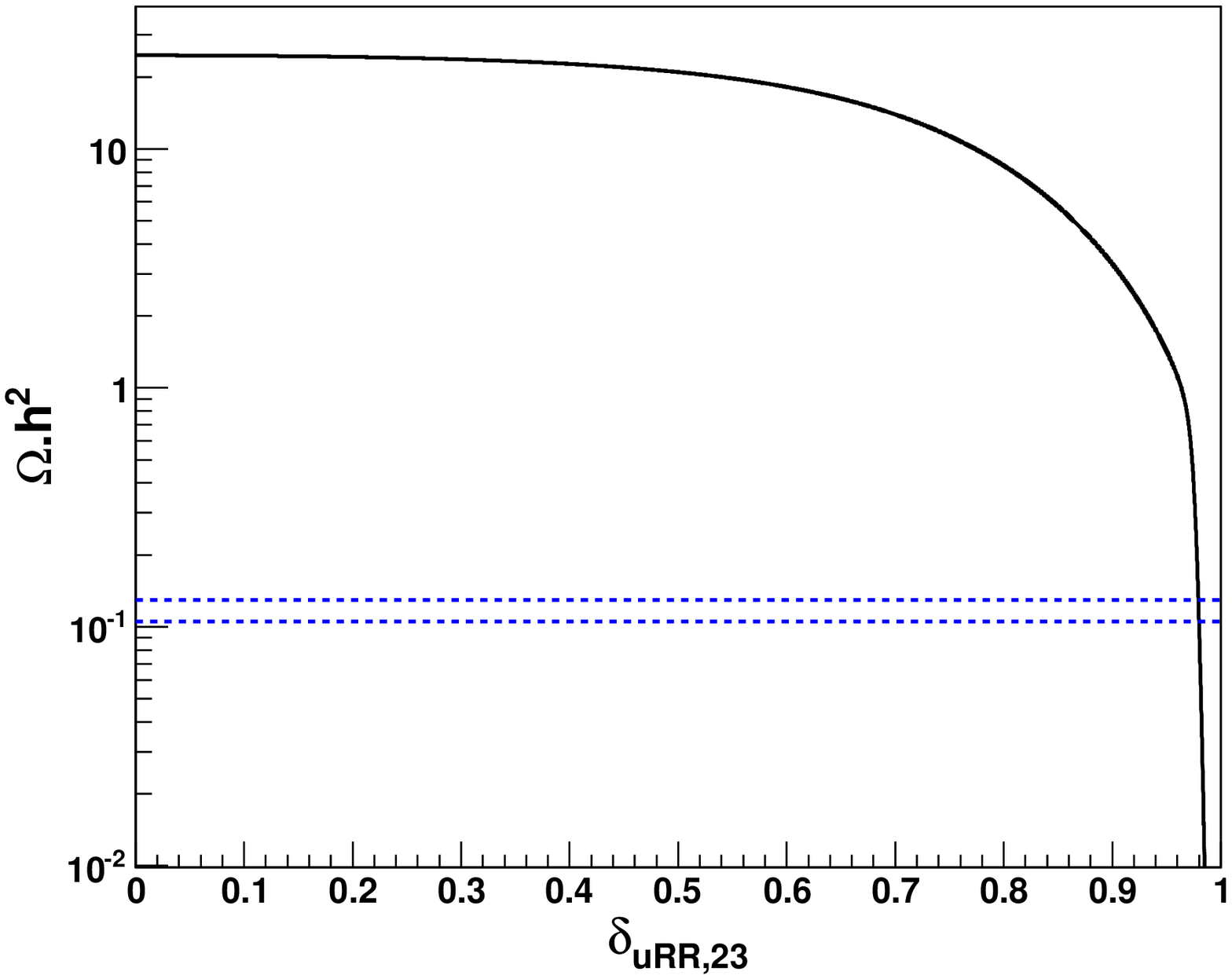}
	\includegraphics[width=0.45\textwidth]{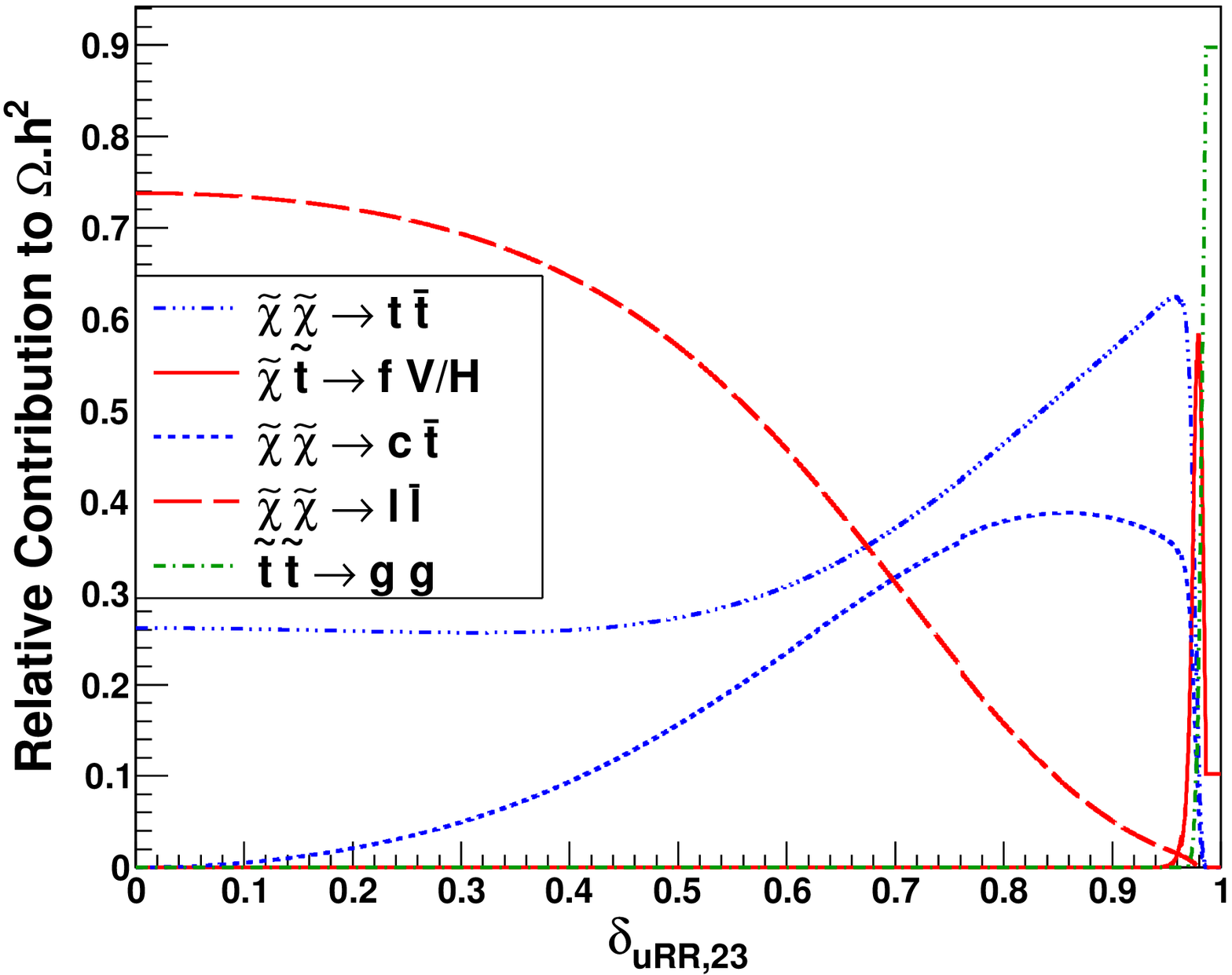}
   \end{center}
   \caption{Relic density of the neutralino (left) and contributing processes (right) as a function of $\delta^{u,\rm RR}_{23}$ for $m_0 = 1500$ GeV, $m_{1/2} = 680$ GeV, $A_0 = -500$ GeV, $\tan\beta = 10$, and $\mu > 0$.}
   \label{fig:CMSSM3}
\end{figure}

In Fig.\ \ref{fig:CMSSM3} we show for a given parameter point the neutralino relic density and the contributing processes as a function of the flavour-violation parameter $\delta^{u,\rm RR}_{23}$. While for the case of MFV, this scenario is cosmologically strongly disfavoured with $\Omega_{\tilde{\chi}_1^0}h^2 \gtrsim 20$, the relic density decreases with increasing flavour mixing to reach the favoured value of $\Omega_{\tilde{\chi}_1^0}h^2 \approx 0.11$ for $\delta^{u,\rm RR}_{23} \sim 0.98$. For low values of $\delta^{u,\rm RR}_{23}$, the annihilation is dominated by lepton final states (about 75\%), which do, however, not lead to a sufficiently enhanced annihilation cross section. The subleading channel is annihilation into top-quark pairs (about 25\%). For $\delta^{u,\rm RR}_{23} \gtrsim 0.2$, flavour violation effects start to manifest by opening the channel $\tilde{\chi}_1^0 \tilde{\chi}_1^0 \to c \bar{t} (t \bar{c})$. The relative contribution of this process amounts to almost 40\% at $\delta^{u,\rm RR}_{23} \sim 0.8$. For $\delta^{u,\rm RR}_{23} > 0.5$, the annihilation into top-quarks is significantly enhanced due to the lighter squark in the $t$-channel propagator, so that this channel remains more important than the newly opened annihilation into top- and charm-quarks. All contributions from neutralino pair annihilation drop at $\delta^{u,\rm RR}_{23} \sim 0.95$ when the squark $\tilde{u}_1$ becomes light enough for efficient coannihilation. The corresponding total relative contribution amounts to about 60\%. When the squark becomes even lighter, also squark pair annihilation into gluon pairs plays an important role (see Eq.\ (\ref{eq:CoAnn})), leading to relative contributions of about 90\% at most.

For this discussed scenario, the favoured relic density of the neutralino is achieved through important coannihilation for rather large values of the flavour mixing parameter $\delta^{u,\rm RR}_{23}$. Note that, depending on the exact parameter point under consideration and the corresponding relic density in the MFV case, this can also happen for lower values of $\delta^{u,\rm RR}_{23}$. In the same way, the enhancement of the total cross section through the new contributions from $c\bar{t} (t\bar{c})$ final states can be sufficient to achieve $\Omega_{\tilde{\chi}^0_1}h^2 \sim 0.11$.

\begin{figure}
   \begin{center}
	\includegraphics[width=0.45\textwidth]{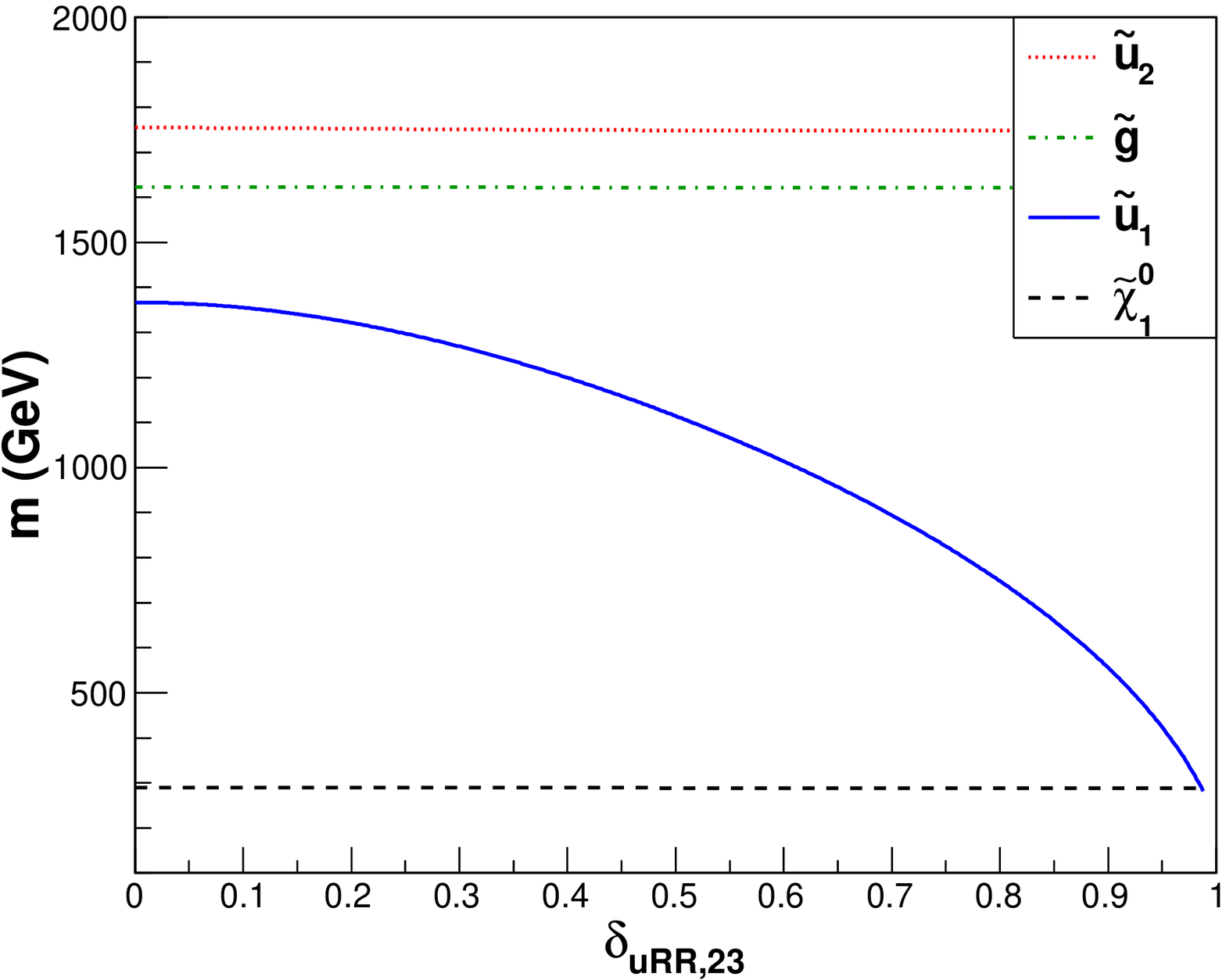}
	\includegraphics[width=0.45\textwidth]{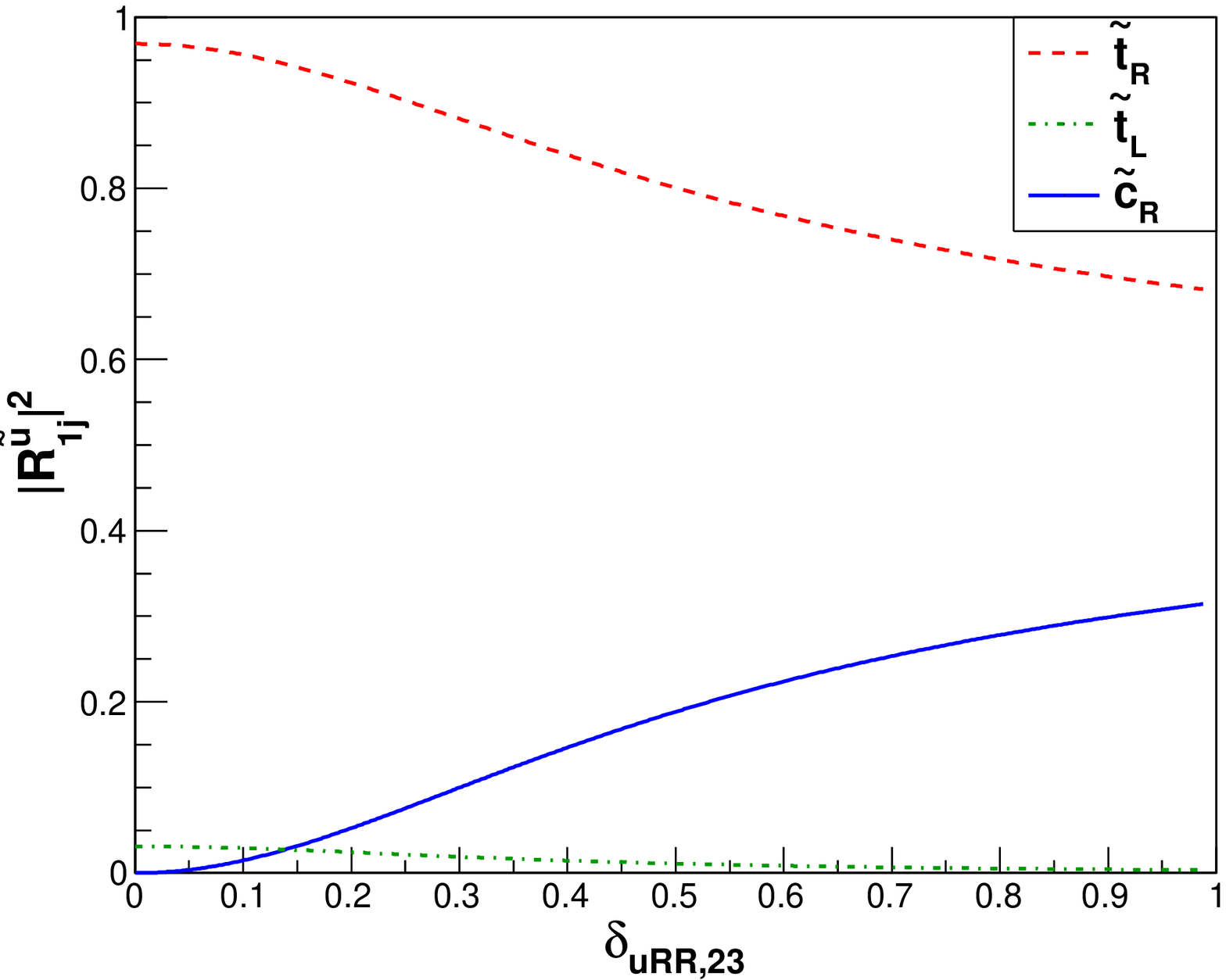}
   \end{center}
   \caption{Masses of the two lightest up-type squarks, gluino, and lightest neutralino (left) and flavour decomposition of lightest up-type squark (right)as a function of $\delta^{u,\rm RR}_{23}$ for $m_0 = 1500$ GeV, $m_{1/2} = 680$ GeV, $A_0 = -500$ GeV, $\tan\beta = 10$, and $\mu > 0$.}
   \label{fig:CMSSM4}
\end{figure}

For completeness, we show in Fig.\ \ref{fig:CMSSM4} the masses of the two lightest up-type squarks, the gluino, and the lightest neutralino as a function of the NMFV-parameter $\delta^{u,\rm RR}_{23}$ as well as the flavour decomposition for the same scenario as discussed above. The squark mass splitting is increased due to the additional off-diagonal entries in the mass matrix, so that the mass of $\tilde{u}_1$ decreases. For large flavour mixing, it comes close to the neutralino mass, leading to the important coannihilation as seen in Fig.\ \ref{fig:CMSSM2}. The masses of $\tilde{u}_2$ ($=\tilde{c}_L$), the neutralino and the gluino remain practically unaffected by the considered generation mixing.

\subsection{Non-universal gaugino masses \label{sec4b}}

When considering $SO(10)$ grand unification theories (GUT), the properties of the SUSY breaking mechanism are related to the breaking of an $SU(5)$ subgroup into the standard model gauge group $SU(3) \times SU(2) \times U(1)$. The relations between the gaugino masses $M_i$ ($i=1,2,3$) at the unification scale are given by the embedding coefficients of the standard model groups in $SU(5)$. In particular, the unification constraint $M_i = m_{1/2}$ of the CMSSM can be relaxed without spoiling the unification of the gauge couplings. Three independent parameters are then needed to fully parameterize the gaugino sector. A possible set is the wino mass $M_2$ together with the two dimensionless variables $x_1 = M_1/M_2$ and $x_3 = M_3/M_2$. The case $x_1=x_3=1$ corresponds to the CMSSM discussed above. Previous studies have shown that non-universal gaugino mass models have an interesting dark matter phenomenology \cite{DM_NUGM, DM_NUGM2}.

\begin{figure}
   \begin{center}
	\includegraphics[width=0.45\textwidth]{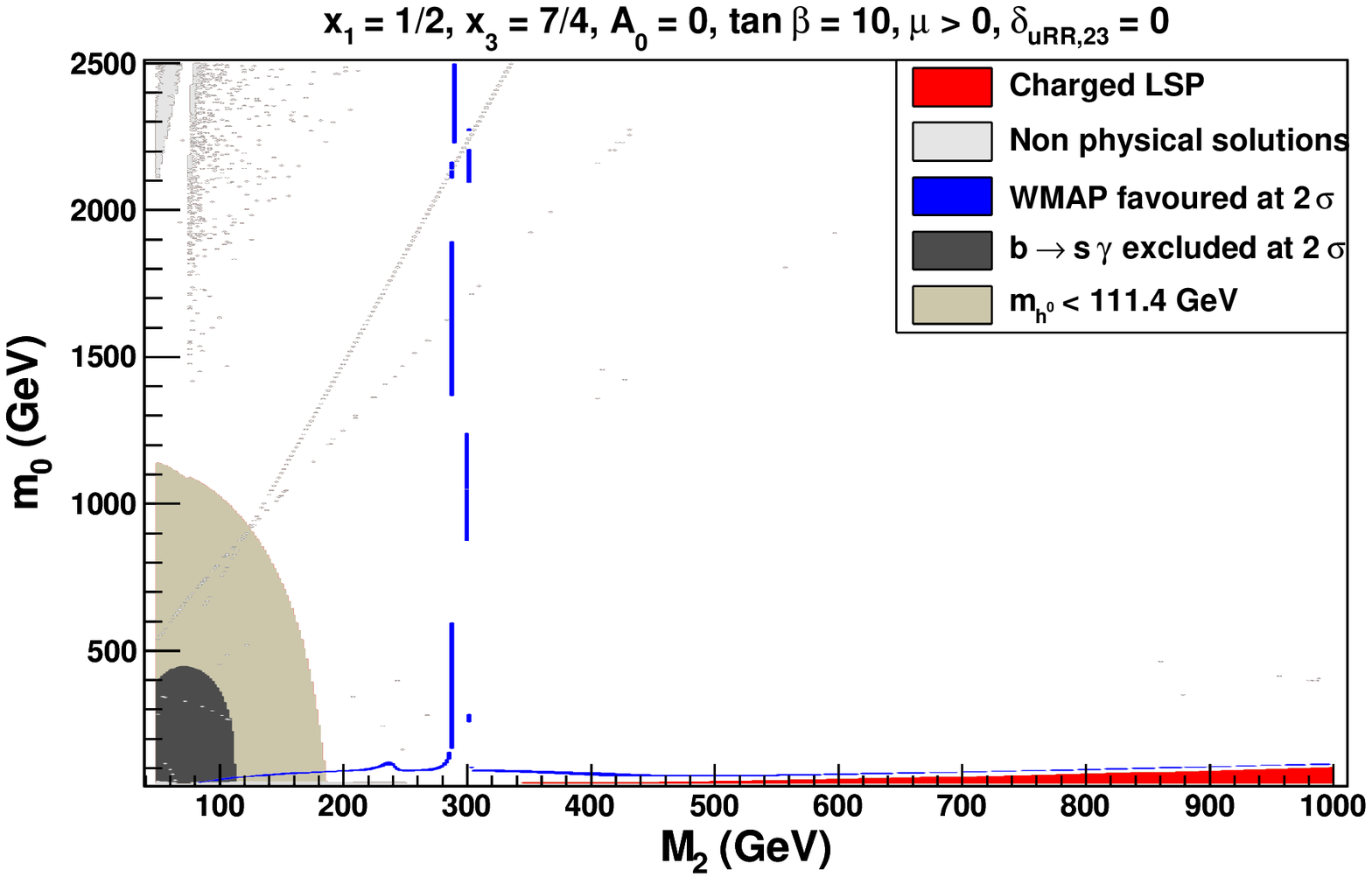}
	\includegraphics[width=0.45\textwidth]{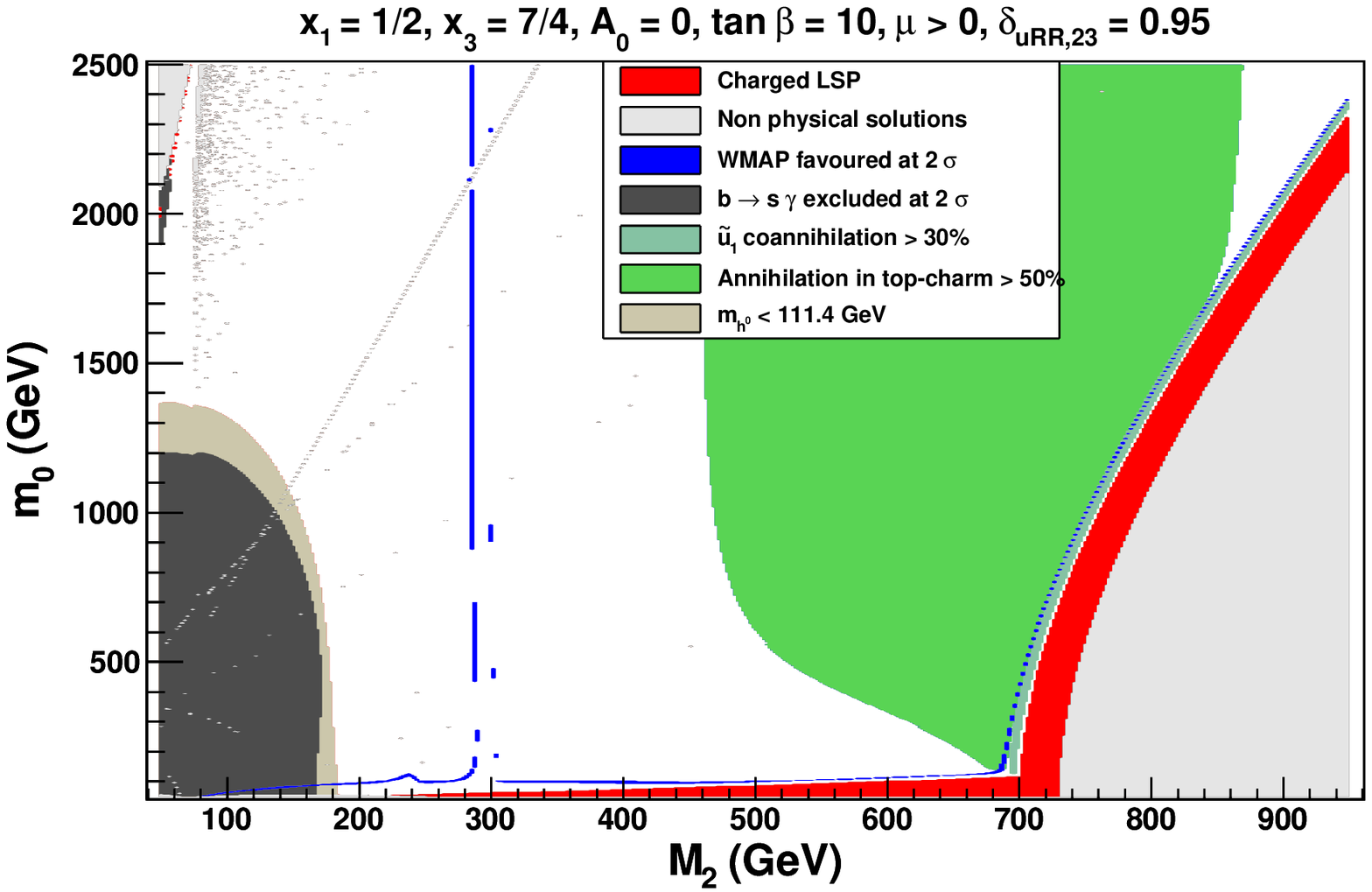}
	\includegraphics[width=0.45\textwidth]{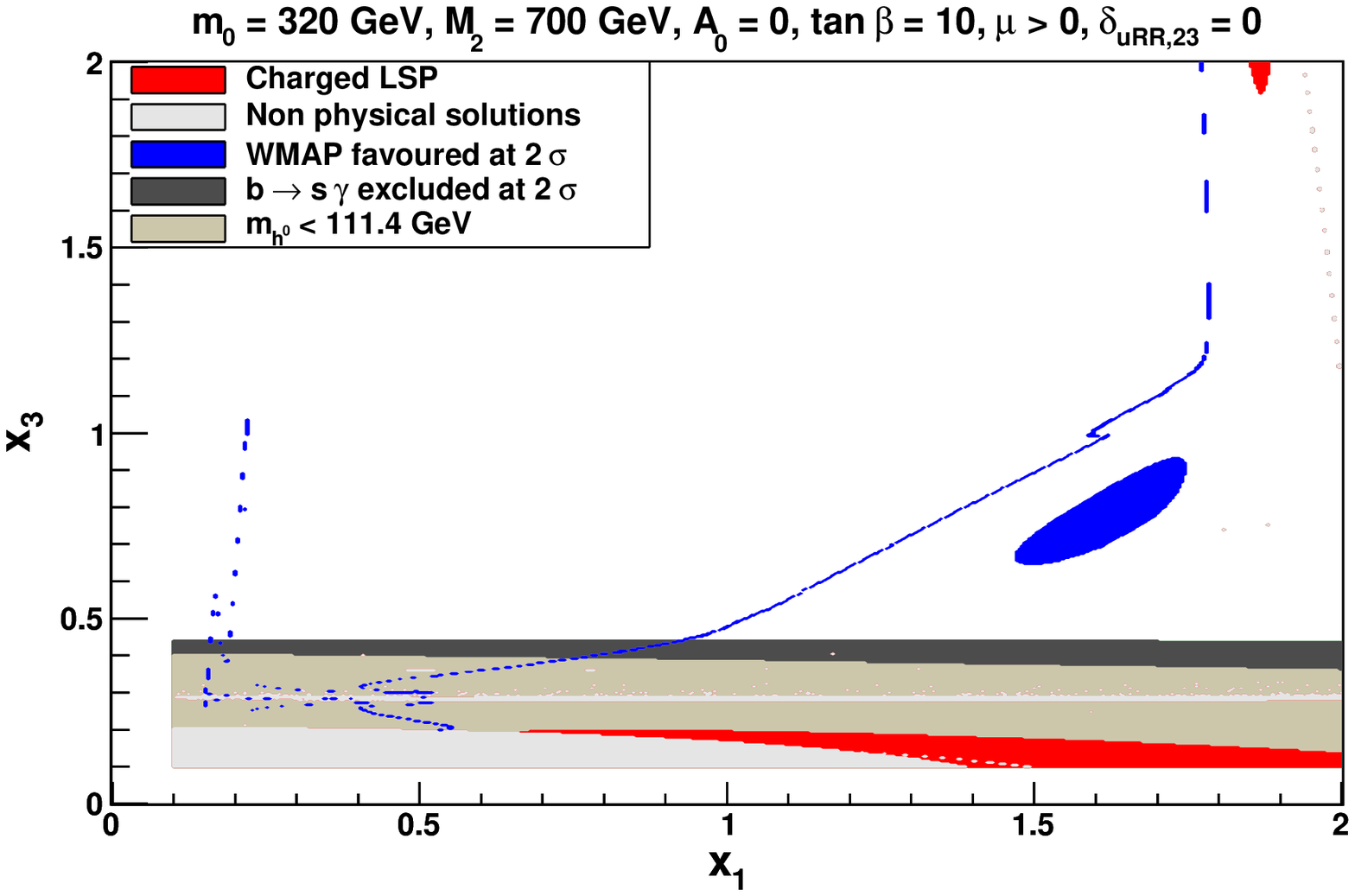}
	\includegraphics[width=0.45\textwidth]{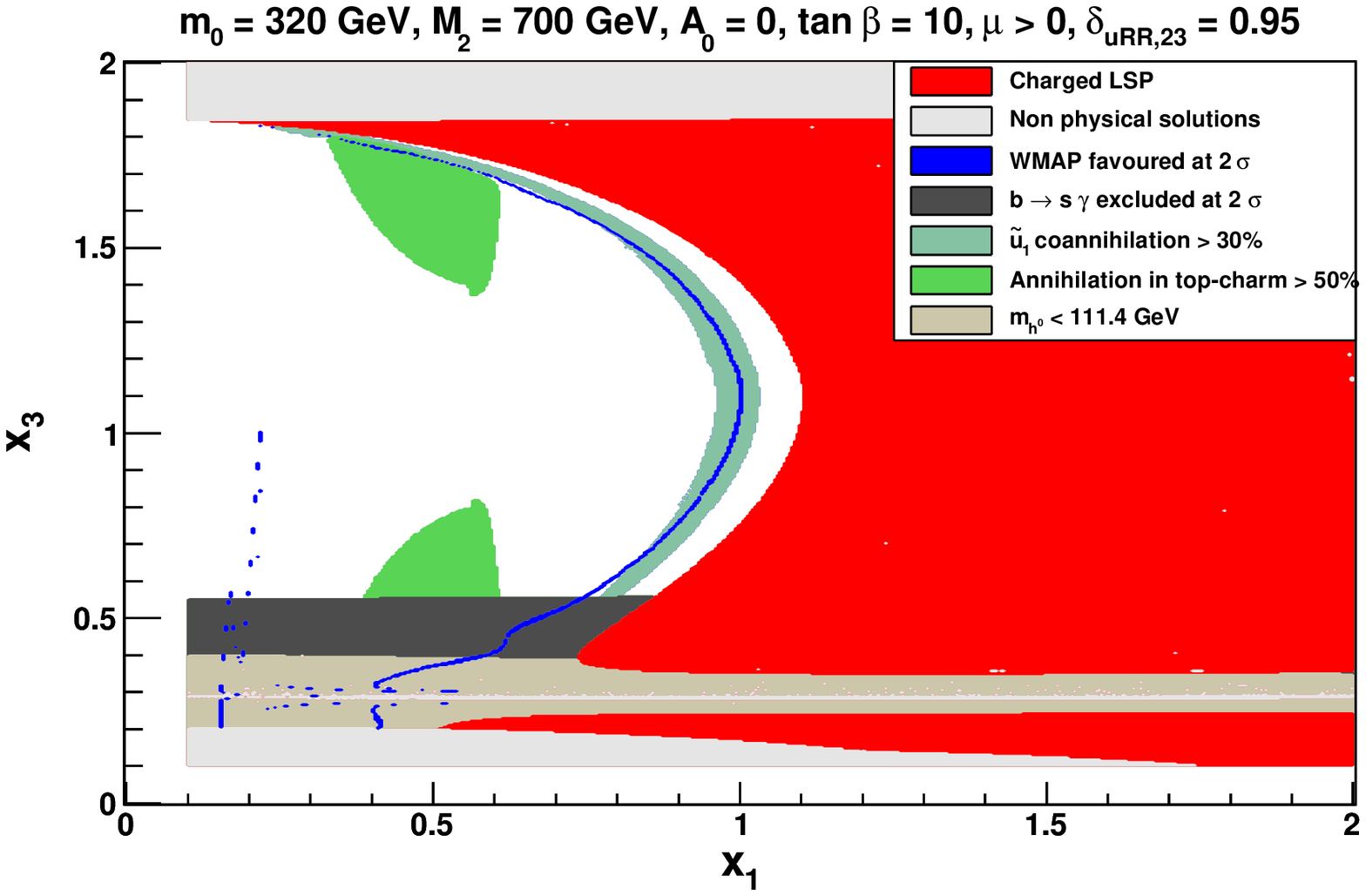}
   \end{center}
   \caption{Top: Constraints in the ($m_0$, $M_{2}$) plane for $x_1 = 1/2$, $x_3 = 7/4$, $\delta^{u,\rm RR}_{23} = 0$ (left) and $\delta^{u,\rm RR}_{23} = 0.95$ (right) in NUGM. Bottom: Constraints in the ($x_{1}$, $x_{3}$) plane for $m_0 = 320$ GeV, $M_{2} = 700$ GeV, $\delta^{u,\rm RR}_{23} = 0$ (left) and $\delta^{u,\rm RR}_{23} = 0.95$ (right).}
   \label{fig:NUGM}
\end{figure}

We start by showing the relevant constraints in the ($m_0$, $M_{2}$) plane for $A_0 = 0$ GeV, $\tan \beta = 10$, $\mu >0$, $x_1 = 1/2$ and $x_3 = 7/4$ in the upper panels of Fig.\ \ref{fig:NUGM}. The flavour violating parameter $\delta^{u,\rm RR}_{23}$ is set to zero (left) and to $0.95$ (right). In both cases, one WMAP-favoured region is situated around the resonance of the lightest Higgs-boson (for $M_{2} \approx 300$ GeV). The neutralino-stau coannihilation region is also present in both cases for $m_0 \approx 100$ GeV, next to the stau-LSP region. For $M_{2} \gtrsim 700$ GeV, due to the large mass splitting for $\delta^{u,\rm RR}_{23} = 0.95$, the squark-LSP region and its neighbouring coannhilation region are present, as it is the case in the CMSSM (see Fig.\ \ref{fig:CMSSM1} right). In this case, a region with sizeable relative contribution from neutralino pair annihilation into $t \bar{c}$ is also present. Note that, due to the non-universality, the neutralino can be lighter as compared to the CMSSM. Therefore, this region is bounded at a certain value of $M_2$, since below this bound the neutralino is not heavy enough to kinematically allow the top production. An upper bound for this region at a higher value of $M_2$ is also observable, since above this value the neutralino is heavy enough to produce top quark pairs. As a result, this region lies in the range $450 \lesssim M_2 \lesssim 850$ GeV. The region excluded by the $b \to s \gamma$ branching ratio is significantly larger for $\delta^{u,\rm RR}_{23} = 0.95$ ($m_0 < 1200$ GeV, $M_{2} < 150$ GeV) than for $\delta^{u,\rm RR}_{23} = 0$ ($m_0 < 450$ GeV, $M_{2} < 100$ GeV). For $\delta^{u,\rm RR}_{23} = 0.95$ this region is similar to the one excluded by the light Higgs mass.

We then show in the lower panels of Fig.\ \ref{fig:NUGM} the constraints in the ($x_1$,$x_3$) plane for a particular point of the parameter space for $\delta^{u,\rm RR}_{23} = 0$ (left) and $\delta^{u,\rm RR}_{23} = 0.95$ (right). Note that the chosen point, for $\delta^{u,\rm RR}_{23} = 0.95$ (i.e right panel) and $x_1 = 1/2$, $x_3 = 7/4$, corresponds to a point where the relic density lies in the WMAP interval (see upper right panel of Fig.\ \ref{fig:NUGM}). For $\delta^{u,\rm RR}_{23} = 0$ different allowed regions are visible. One is the very low $x_1$ ($\approx 0.2$) region in which the neutralino annihilates mainly to pairs of $b$ quarks or tau leptons via a light Higgs resonance. Then there is a diagonal line corresponding to the heavy Higgs resonance (the neutralino mass increases with $x_1$ and the heavy Higgs mass with $x_3$). For large $x_1$ ($\gtrsim 1.8$) the neutralino becomes mainly wino and annihilates strongly into $W$ boson pairs, thus decreasing the relic density below the lower limit. In the favoured ellipse-shaped region for $x_1 \simeq 1.6$, $x_3 \simeq 0.8$, the heavy Higgs resonance and $W$ boson final states processes contribute in a way that the relic density is compatible with the required interval. For $\delta^{u,\rm RR}_{23} = 0.95$ it is immediately visible that new allowed regions appear. In the low $x_3$ region annihilation of neutralinos into $t\bar{c}$ (and $t\bar{t}$ for $x_1 \gtrsim 0.6$) are the main contributions. However these new contributions essentially do not modify the shape of the allowed region compared to the $\delta^{u,\rm RR}_{23} = 0$ case. Moreover, we can notice that this region is excluded by BR$(b \to s \gamma)$. The interesting region is for larger $x_3$, where the neutralino coannihilation with the lightest squark appears. In this region the mass difference between the neutralino and the lightest squark is approximatively $30$ GeV and the main contributions are neutralino-squark annihilation into top/charm quark and gluon, and squark annihilation into gluon pairs. While the neutralino mass increases with $x_1$, the lightest squark mass increases with $x_3$ until some point, and then decreases, which explains the particular arc shape of this region. Therefore this coannihilation region is relatively symmetric with respect to $x_3 \approx 1.1$, and for each favoured point with $x_3 < 1.1$ there is a corresponding favoured point with $x_3 > 1.1$. As explained below, the $x_3 < 1.1$ part of this region is more strongly constrained by BR$(b \to s \gamma)$, which explains our choice for the $x_3$ value in the upper plots. 

Compared to the CMSSM, the predicted value for BR$(b \to s \gamma)$ is much less constraining. First, because of the chosen value of $A_0$, but also because of the spectrum modification induced by the non-universal gaugino masses. This is illustrated in the lower right panel of Fig.\ \ref{fig:NUGM}, where $m_0$ and $M_2$ have been fixed and the (dis)favoured regions are shown in the ($x_1$,$x_3$) plane. It is striking that BR$(b \to s \gamma)$ has a strong dependence on $x_3$, which influences the whole SUSY spectrum through the renormalization group running. Most of the non standard model contribution to BR$(b \to s \gamma)$ comes from chargino and charged Higgs loops, the former being negative, and the latter positive. At low $x_3$, the (negative) chargino contribution is dominant, which leads to a branching ratio far below the standard model prediction. With increasing $x_3$ sparticles become heavier, and all absolute values of contributions decrease. However the absolute value of the chargino contribution come closer to the one from Higgs bosons, which leads to important cancellations and the branching ratio gets very close to the standard model value. For large $x_3$, the chargino contribution (in absolute value) becomes smaller than the Higgs contribution. The sum is then positive, but rather small as the masses are large.

Among the other observables given in Tab.\ \ref{tab1}, the ones which exclude some regions of the parameter space presented here are $\Delta a_{\mu}$ and $\Delta \rho$, but only in the very low mass region which is already partially excluded by $\textnormal{BR}(b \to s \gamma)$. For the plots in the ($m_0$, $M_{2}$) plane the neutralino mass excludes the region $M_{2} < 250$ GeV and the chargino mass excludes the region $M_{2} < 150$ GeV. The stop mass excludes also some region for which it is the LSP (i.e.\ already excluded). For the plots in the ($x_{1}$, $x_{3}$) plane, the neutralino mass excludes the region $x_1 < 0.1$.

\subsection{Non-universal Higgs masses \label{sec4c}}

Similarly to the mechanism leading to non-universal gaugino masses in $SO(10)$ SUSY GUTs, depending on the exact representation to which the Higgs doublets belong, their corresponding SUSY breaking masses $m_{H_D}$ and $m_{H_U}$ need not necessarily be the same. In non-universal Higgs mass models they can therefore be treated as independent parameters at the high scale \cite{DM_NUHM2, DM_NUHM3}. For $m_{H_U} = m_{H_D} = m_0$ the standard CMSSM is recovered.

\begin{figure}
   \begin{center}
	\includegraphics[width=0.45\textwidth]{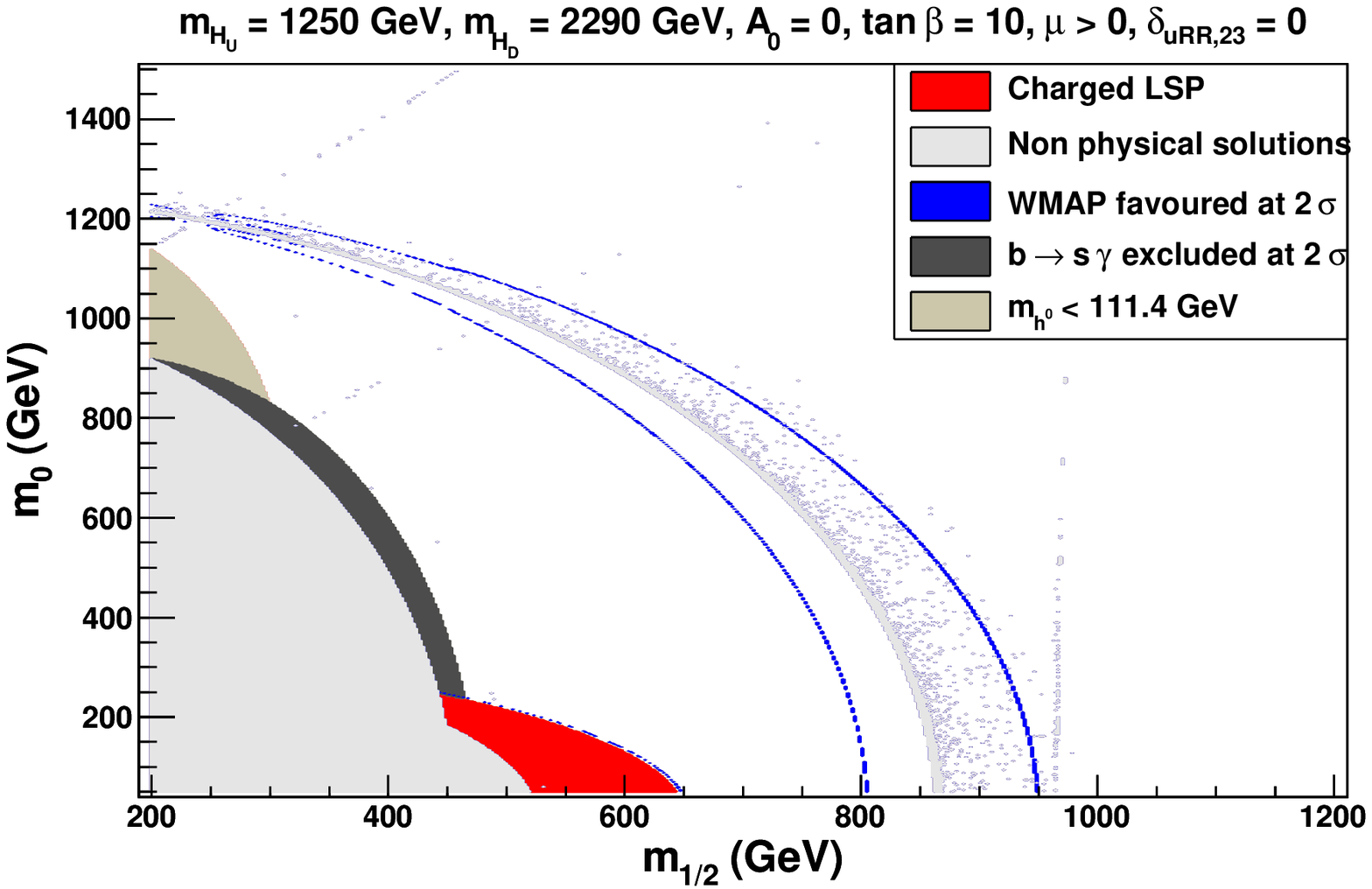}
	\includegraphics[width=0.45\textwidth]{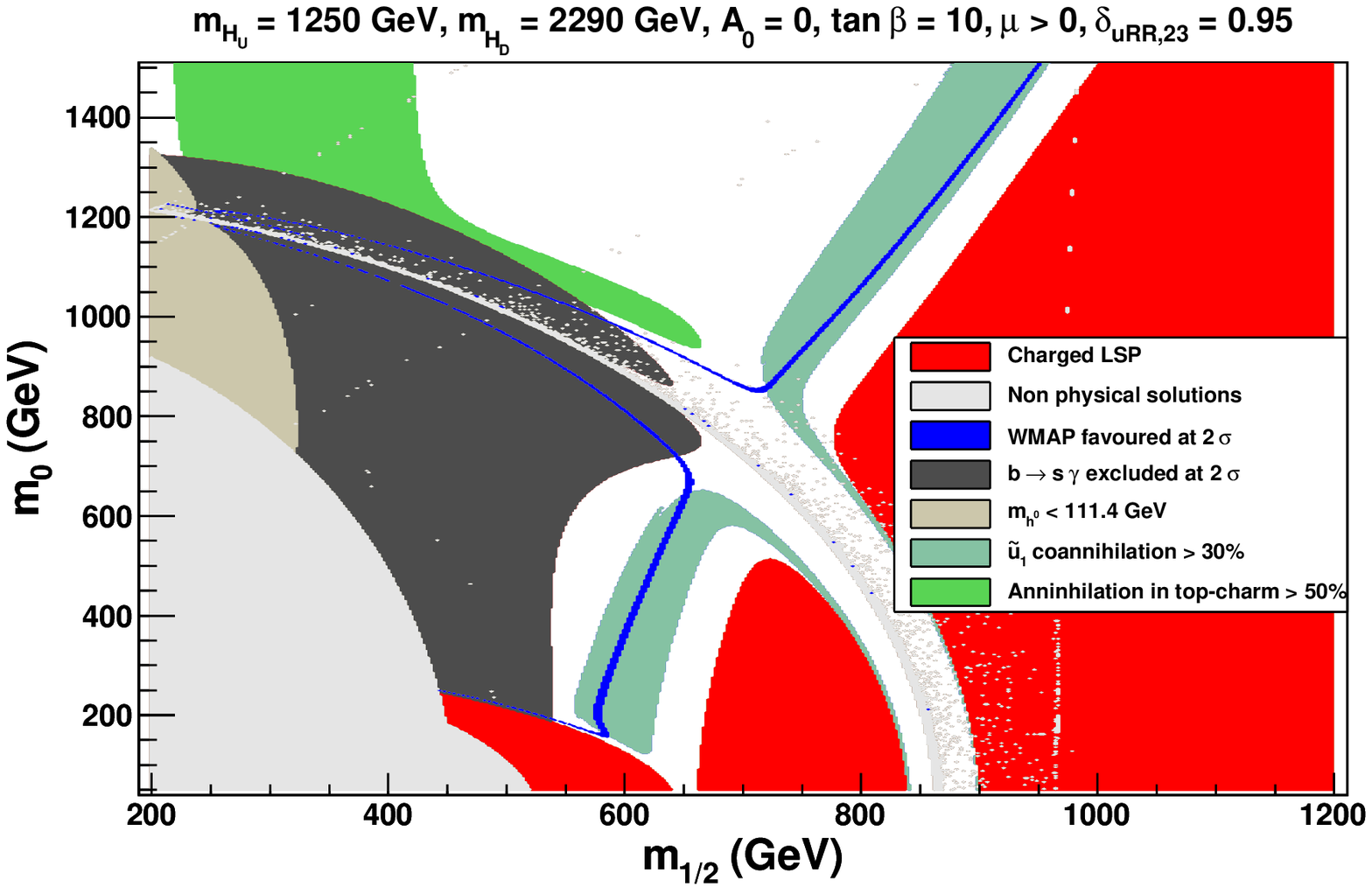}
	\includegraphics[width=0.45\textwidth]{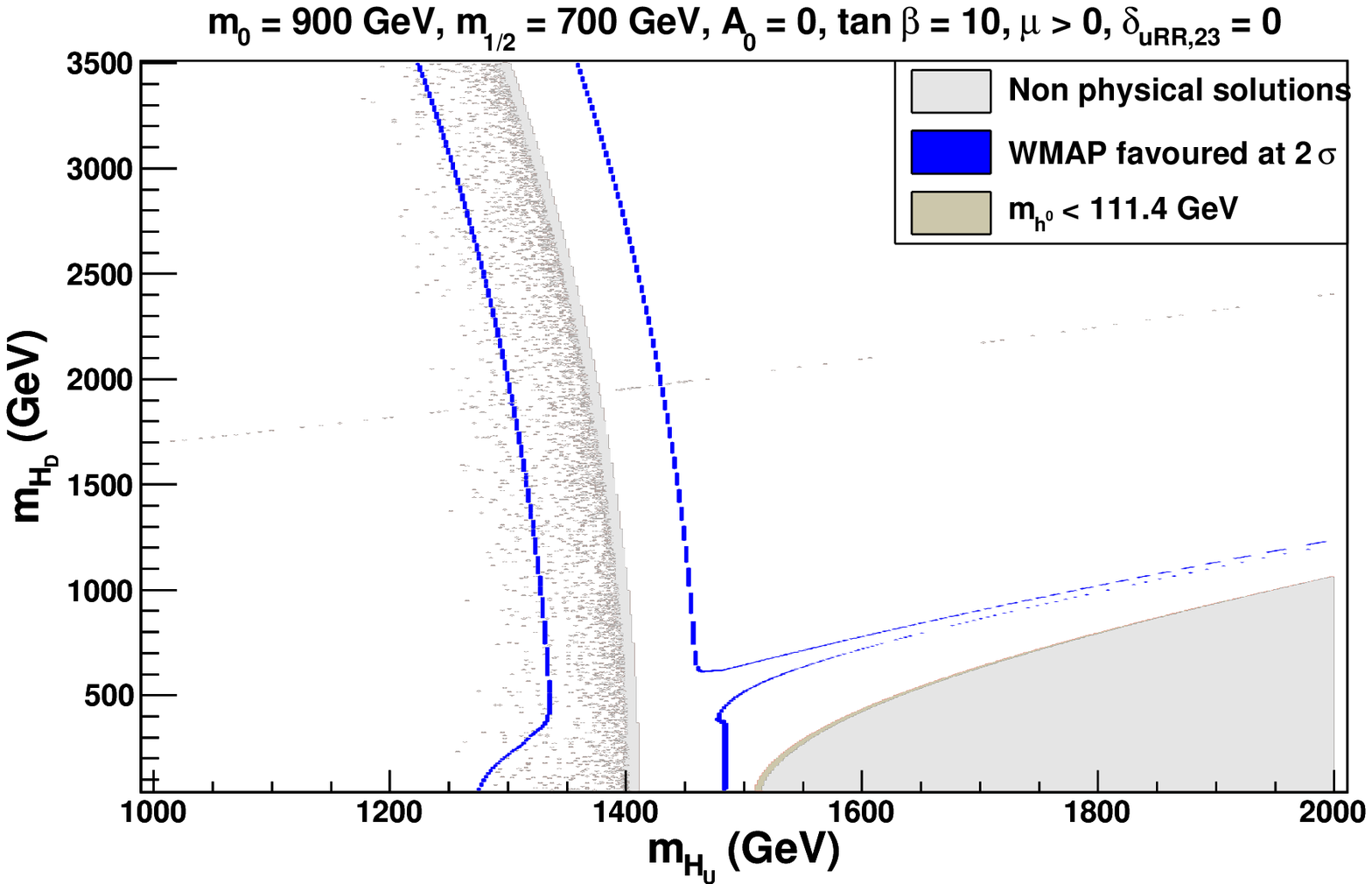}
	\includegraphics[width=0.45\textwidth]{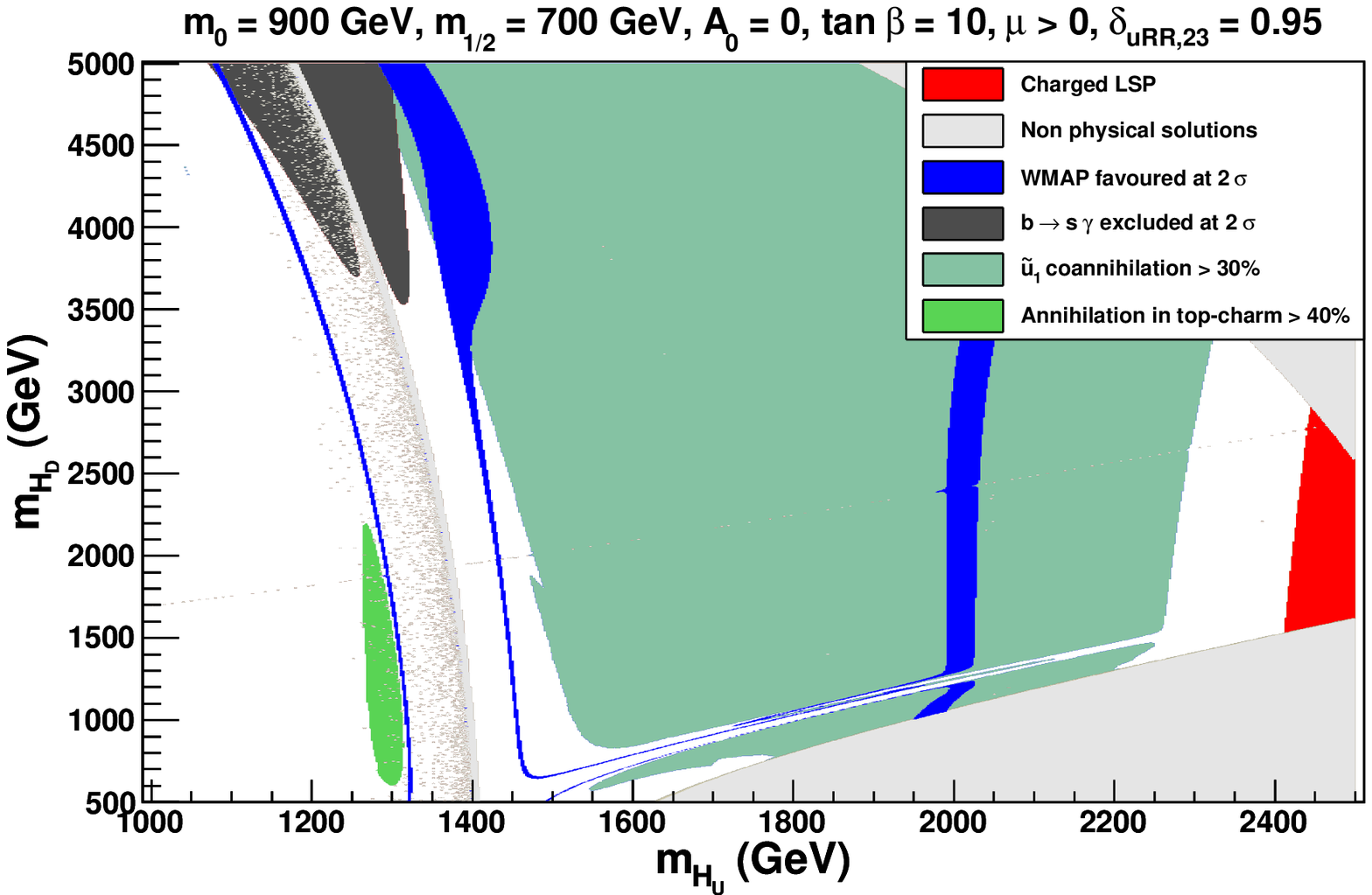}
   \end{center}
   \caption{Top: Constraints in the ($m_0$, $m_{1/2}$) plane for $\delta^{u,\rm RR}_{23} = 0$ (left) and $\delta^{u,\rm RR}_{23} = 0.95$ (right) in NUHM. Bottom: Constraints in the ($m_{H_U}$, $m_{H_D}$) plane for $\delta^{u,\rm RR}_{23} = 0$ (left) and $\delta^{u,\rm RR}_{23} = 0.95$ (right).}
   \label{fig:NUHM}
\end{figure}

We start by studying the ($m_0$, $m_{1/2}$) plane for fixed $m_{H_U} = 1250$ GeV, $m_{H_D} = 2290$ GeV, $A_0=0$, $\tan\beta=10$, and $\mu>0$. The resulting excluded and cosmologically favoured regions are shown in Fig.\ \ref{fig:NUHM} (upper panels), again for both the case of MFV ($\delta^{u,\rm RR}_{23}=0$) and a rather large flavour mixing parameter $\delta^{u,\rm RR}_{23}=0.95$. For $\delta^{u,\rm RR}_{23} = 0$, the only allowed regions arise from the coannihilation of the neutralino with the superpartners of the tau or neutrinos and from the annihilation of neutralino pairs into $W^{\pm}$-bosons and top quark pairs due to the high higgsino component. The latter region is actually divided into two parts, parallel to the excluded region where certain squared sfermion masses become negative. 

For $\delta^{u,\rm RR}_{23} = 0.95$, new contributions from $\tilde{\chi}_1^0 \tilde{\chi}_1^0 \to c \bar{t} (t \bar{c})$ make their appearance, as it is also the case in the discussed CMSSM and NUGM scenarios. Here, however, important coannihilations of the neutralino with the lightest squark are present, leading to a completely modified picture with respect to the MFV case. Among the WMAP favoured regions, only the one due to coannihilation survives the $b \to s \gamma$ constraint. The discussion of this constraint (as all the constraints given in Tab.\ \ref{tab1}) is here similar to the CMSSM one.

Let us now study the ($m_{H_U}$, $m_{H_D}$) plane for fixed values of $m_0=900$ GeV and $m_{1/2}=700$ GeV, shown in Fig.\ \ref{fig:NUHM} (lower panels). For $\delta^{u,\rm RR}_{23} = 0$ the situation is quite similar as described above. Two parallel allowed regions, where the neutralino is strongly higgsino and annihilating through the light Higgs resonance, are present. In addition, two other allowed regions, corresponding to the heavy neutral Higgs resonance, are observed for large $M_{H_U}$. Allowing for flavour violation ($\delta^{u,\rm RR}_{23} = 0.95$), two large additional allowed regions appear, where neutralino-squark coannihilation processes are dominant (up to $50 \%$ including light squark annihilation into gluons). Note that the corresponding WMAP-favoured areas are very large as compared to $\delta^{u,\rm RR}_{23} = 0$. The flavour violating annihilation channel $\tilde{\chi}_1^0 \tilde{\chi}_1^0 \to c \bar{t} (t \bar{c})$ is less important in this case.

Again, all constraints have been checked to be fulfilled in the shown parameter space, except for the light neutralino/chargino mass in the region which lies close to the unphysical region.

\section{LHC phenomenology \label{sec5}}

Finally, we discuss the collider phenomenology corresponding to CMSSM scenarios that feature new annihilation or coannihilation channels induced through flavour violating elements. Typical signatures for quark flavour violation in the context of squark production at hadron colliders have been discussed in Refs.\ \cite{HurthPorod, NMFV_Squark1, NMFV_Squark2}. A particularly promising process is the production of the lightest squark-antisquark pair, and their subsequent decay into charm- and top-quarks. The rather clean signature $pp \to \tilde{u}_1 \tilde{u}_1^* \to c \bar{t} (t \bar{c}) \tilde{\chi}_1^0 \tilde{\chi}_1^0$, where the neutralinos would manifest as sizeable missing energy, might lead to up to $10^4$ events at the LHC with $\sqrt{s}=14$ TeV and an integrated luminosity of 100 fb$^{-1}$ \cite{NMFV_Squark2}. 

Alternatively, new contributions to the decay $t\to c\gamma$ can increase its branching ratio to as much as 10$^{-6}$ and thus render it detectable, e.g., in $t\bar{t}$ production at the LHC, as has been pointed out in Refs.\ \cite{deDivitiis1997, Delepine2004}. In the case in which $\delta_{23}^{u,LR}$ is strongly constrained, e.g.\ by the $B_s$ mixing, and only $\delta_{23}^{u,RR}$ is large, several years of high-luminosity operation might, however, be required.

In order to evaluate the production of squarks and gluinos at the LHC, we have computed the relevant cross sections using the Monte-Carlo package {\tt WHIZARD 1.95} together with the associated matrix element generator {\tt O'MEGA} \cite{WhizardOmega}, where the MSSM with the most general generation mixing as discussed in Sec.\ \ref{sec2} has been implemented \cite{NMFV_Squark2}. We have employed the {\tt CTEQ6L} \cite{CTEQ} set for the parton distribution functions, the factorization scale being set to the average of the produced masses. Finally, the branching ratios of squarks and gluinos have been obtained using {\tt SPheno} \cite{SPheno}. 

\begin{figure}
	\includegraphics[width=0.45\textwidth]{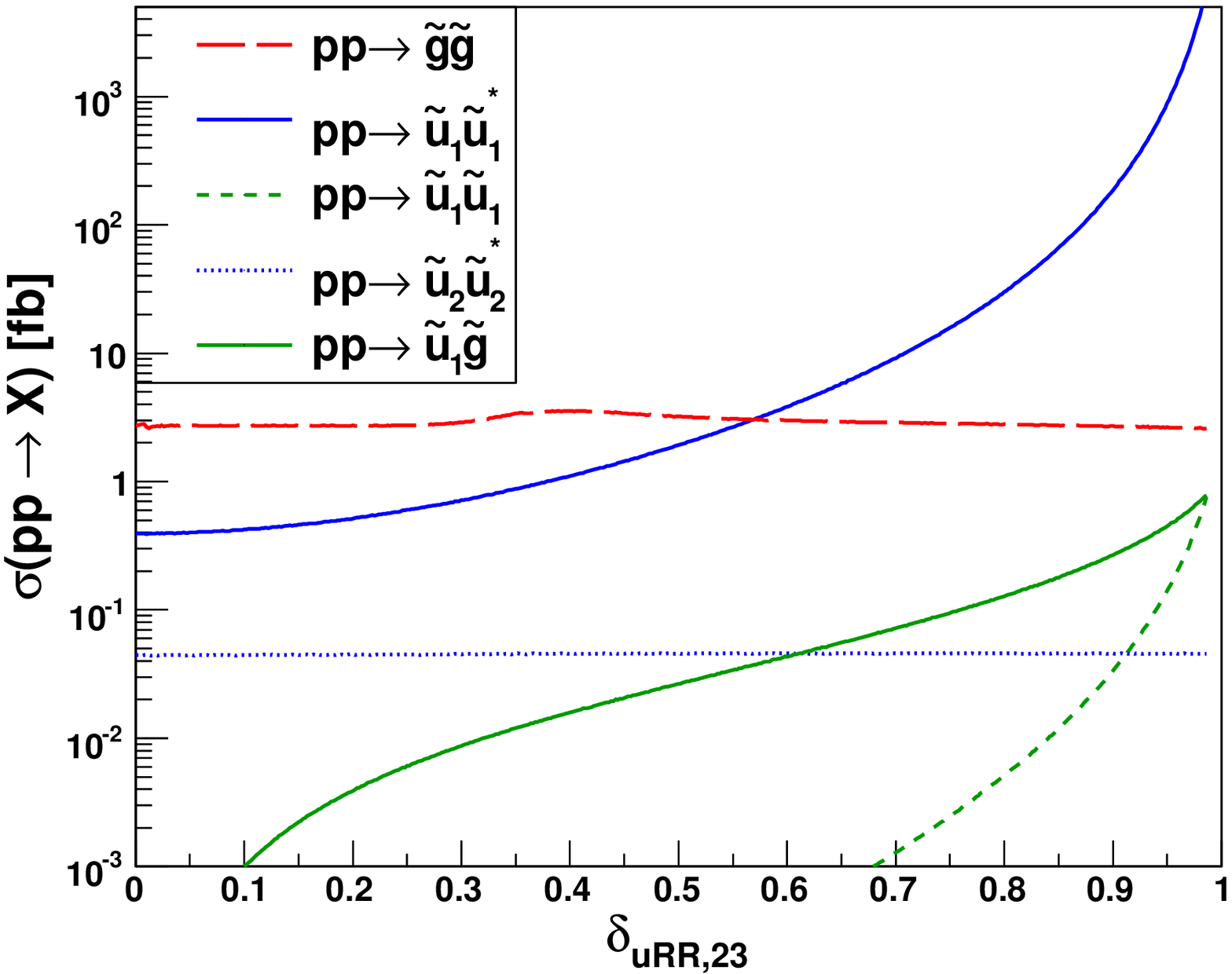}	
	\includegraphics[width=0.45\textwidth]{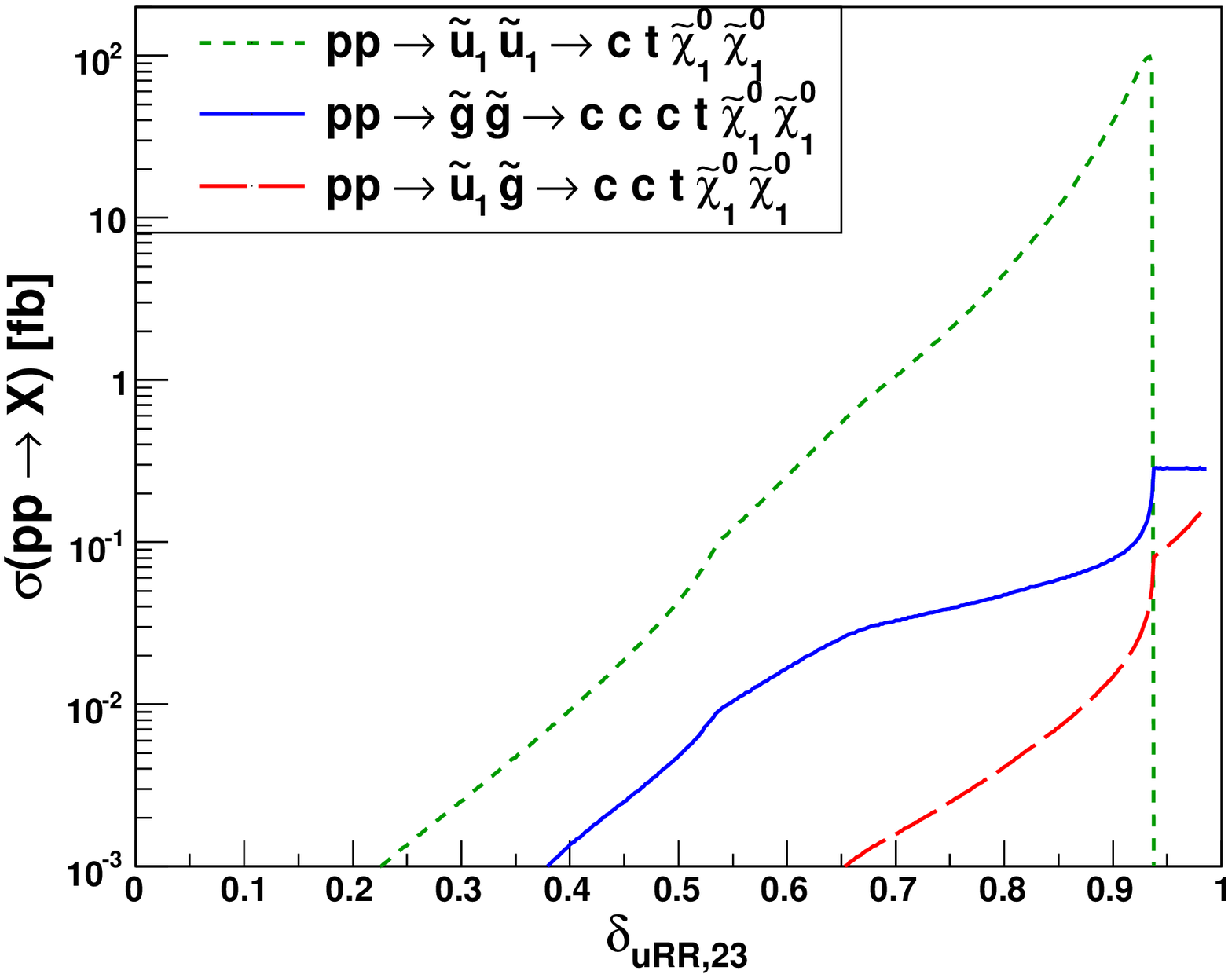}
	\caption{Dominant production cross-sections of up-type squarks and gluinos (left) and resulting NMFV signal cross-section (right) for the CMSSM scenario $m_0=1500$ GeV, $m_{1/2}=680$ GeV, $A_0 = -500$ GeV, $\tan\beta=10$, and $\mu>0$ as a function of the NMFV-parameter $\delta^{u,\rm RR}_{23}$ at the LHC with $\sqrt{s}=14$ TeV. The notation in the legend of the right panel is according to Eqs.\ (\ref{eq:signal0}) and (\ref{eq:signal2}), taking into account all possible combinations of (s)quarks and anti(s)quarks.}
	\label{fig:LHC}
\end{figure}

In the first graph of Fig.\ \ref{fig:LHC}, we show the obtained dominant production modes of squarks and gluinos at the LHC with $\sqrt{s}=14$ TeV for the example scenario already discussed in Sec.\ \ref{sec4a}. In the case of MFV ($\delta^{u,\rm RR}_{23}=0$), gluino pair production is dominant due to the colour structure, while squark-antisquark production is the subdominant channel. Due to the lighter mass, production of $\tilde{u}_1$ is preferred over the production of $\tilde{u}_2$. We focus therefore on the production and decay of the lightest squark and the gluino.

Due to only flavour-diagonal couplings, the gluino pair production cross section is practically independent of the NMFV-parameter $\delta^{u,\rm RR}_{23}$. Contrary, the squark-antisquark pair production receives new contributions in a similar way as the neutralino pair annihilation discussed in Sec.\ \ref{sec3}. Since $\tilde{u}_1$ now has a sizeable $\tilde{c}$-admixture, initial states containing charm-quarks can now contribute. Moreover, the lighter squark in the $t$-channel propagator enhances this channel. Finally, also the phase space is increased due to the decreased squark mass. Taking into account all these effects, the production of $\tilde{u}_1\tilde{u}_1^*$ becomes the dominant channel for $\delta^{u,\rm RR}_{23} \gtrsim 0.5$ and reaches production cross sections of up to $10^3$ fb for $\delta^{u,\rm RR}_{23} \sim 0.95$. For lower mixing parameters, gluino pair production remains numerically most important. 

Similar arguments hold for the associated production of a gluino and a squark. For the same reasons as given above, the production of $\tilde{u}_1\tilde{g}$ is enhanced for large flavour mixing as compared to the MFV case. Note that, although the charge conjugated channel $pp \to \tilde{g}\tilde{u}_1^*$ is not shown in Fig.\ \ref{fig:LHC} (left), it is taken into account in the following calculation of event rates. The associated production of $\tilde{g}$ and heavier squarks is negligible in this context. The practically only flavour content of the second lightest squark $\tilde{u}_2$ is $\tilde{t}_L$, such that its production remains insensitive to the discussed flavour mixing in the right-right sector.

\begin{figure}
	\includegraphics[width=0.45\textwidth]{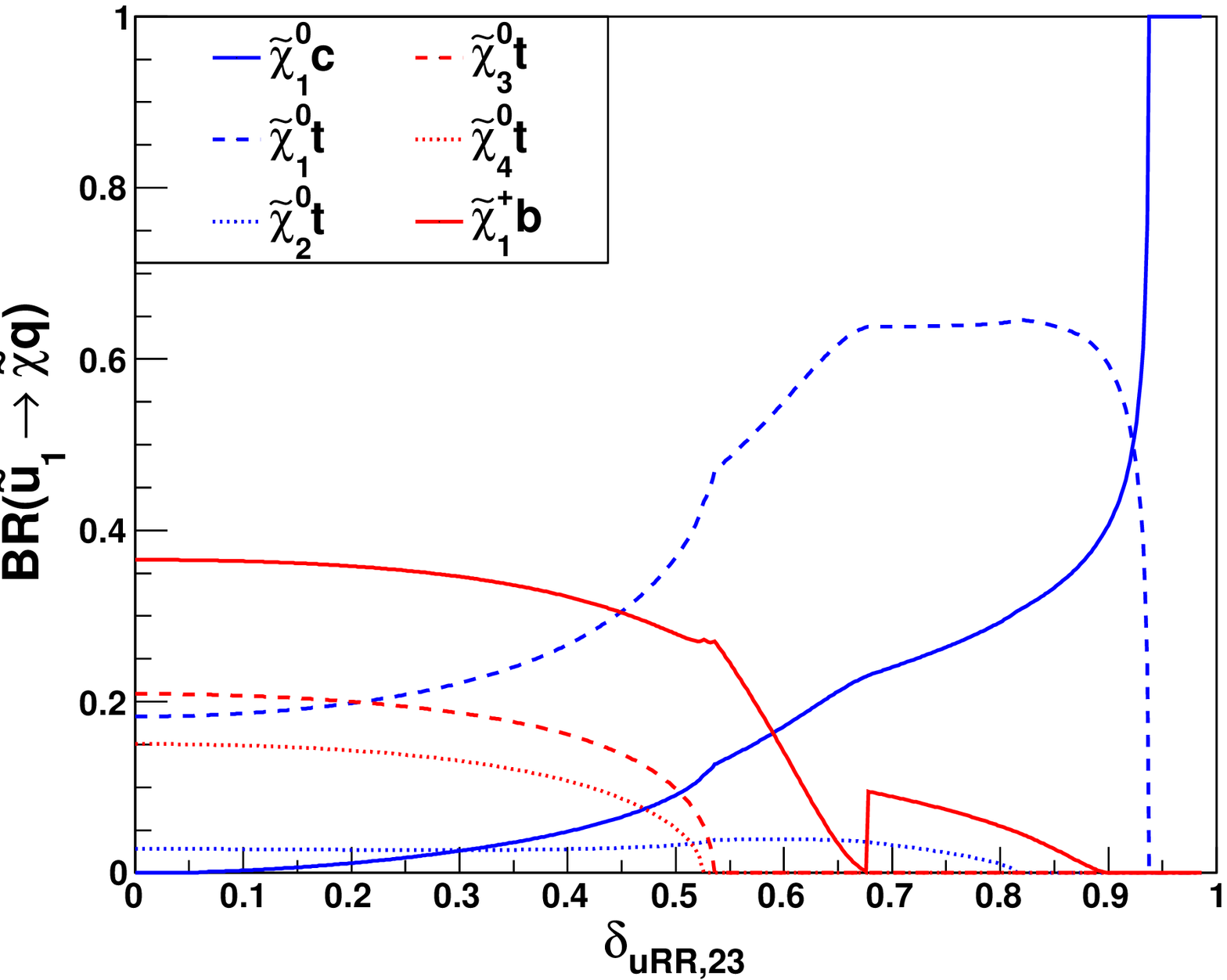}
	\includegraphics[width=0.45\textwidth]{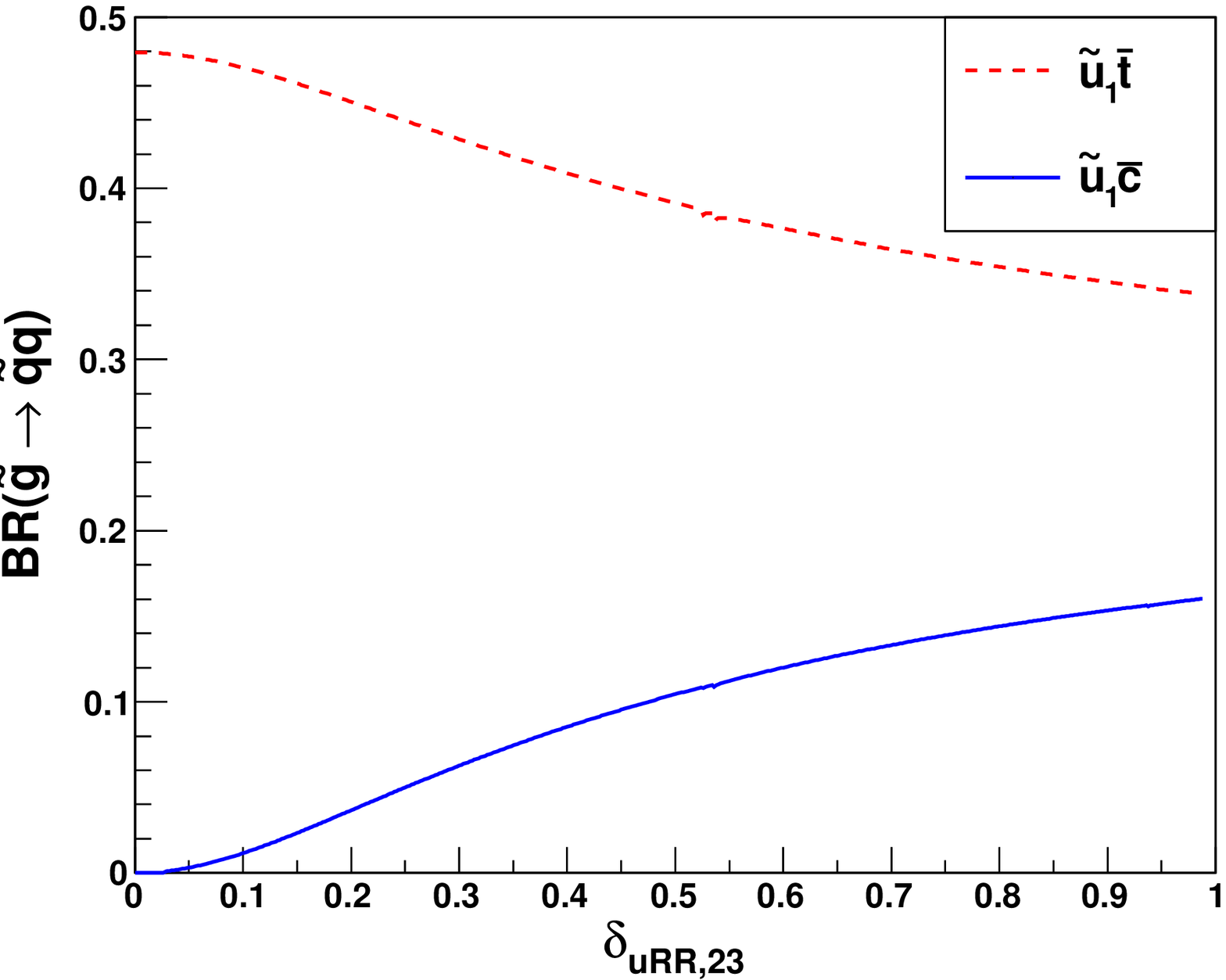}
	\caption{Branching ratios of the lightest up-type squark (left) and the gluino (right) for the scenario of Fig.\ \ref{fig:LHC} as a function of the NMFV-parameter $\delta^{u,\rm RR}_{23}$.}
	\label{fig:BR}
\end{figure}

The branching ratios of squarks and gluinos are also affected by flavour-violating elements, as can be seen from Fig.\ \ref{fig:BR}. Since the lightest squark $\tilde{u}_1$ is a pure stop-like state in the case of MFV, it dominantly decays into $\tilde{\chi}^+_1 b$ and $\tilde{\chi}^0_i t$ ($i=1,2,3,4$). Note that the decay into $\tilde{\chi}_2^0$ is relatively small, since $\tilde{\chi}_2^0$ is almost purely wino-like and couples only to the tiny $\tilde{t}_L$-component of $\tilde{u}_1$. For increasing $\delta^{u,\rm RR}_{23}$, the $\tilde{c}_R$-content increases, and decays into final states including second generation quarks open up. Decays into heavier neutralinos are kinematically forbidden for $\delta^{u,\rm RR}_{23} \gtrsim 0.5$. Although $\tilde{t}_R$ remains the dominant flavour in $\tilde{u}_1$ (see Fig.\ \ref{fig:CMSSM4}), its decay into $\tilde{\chi}_1^0 t$ is closed for $\delta^{u,\rm RR}_{23} \gtrsim 0.9$, since the mass difference between squark and neutralino is smaller than the top mass (see also Fig.\ \ref{fig:CMSSM4}). The only remaining decay mode is then $\tilde{u}_1 \to \tilde{\chi}_1^0 c$. 

As a consequence, the NMFV signature 
\begin{equation}
	pp \to \tilde{u}_1 \tilde{u}_1^* \to c \bar{t}\,(t \bar{c})\,\tilde{\chi}_1^0 \tilde{\chi}_1^0
\label{eq:signal0}
\end{equation} 
discussed in Ref.\ \cite{NMFV_Squark2} cannot be realized for large $\delta^{u,\rm RR}_{23} \gtrsim 0.9$. This channel can have sizeable event rates only for smaller flavour mixing parameters. This can be seen in the second graph of Fig.\ \ref{fig:LHC}, where we show the production cross section combined with the relevant branching ratios in order to estimate the signal rate at the LHC. 

For such important flavour mixing, a moderately sizeable signal rate can, however, stem from gluino pair and from gluino-squark production. In the former case, NMFV final states can be achieved through 
\begin{equation}
	pp \to \tilde{g} \tilde{g} \to c \tilde{u}_1 \, t \tilde{u}_1 \to c c c t \,\tilde{\chi}^0_1 \tilde{\chi}^0_1, 
\label{eq:signal1}
\end{equation} 
where one of the produced gluinos decays into a top quark. As can be seen in Fig.\ \ref{fig:BR}, this decay mode remains allowed (and even dominant) for all values of $\delta^{u,\rm RR}_{23}$. The notation of the final state in Eq.\ (\ref{eq:signal1}) is understood to include all possible combinations of quarks and antiquarks.

The second possibility, mediated through associated production of a squark and a gluino, gives rise to signal events of the type
\begin{equation}
	pp \to \tilde{u}_1 \tilde{g} \to c \tilde{\chi}^0_1 \,t \tilde{u}_1 \to c c t \,\tilde{\chi}^0_1 \tilde{\chi}^0_1, 
\label{eq:signal2}
\end{equation} 
where again the gluino decays into a top quark and the process is understood to include all possible combinations of (s)quarks and anti(s)quarks. 

In the second graph of Fig.\ \ref{fig:LHC}, we show the mentioned signal cross sections as a function of the NMFV-parameter $\delta^{u,\rm RR}_{23}$. As discussed above, the signature in Eq.\ (\ref{eq:signal0}) increases with $\delta^{u,\rm RR}_{23}$, but drops when $m_{\tilde{u}_1} - m_{\tilde{\chi}_1^0} < m_{\rm top}$, i.e.\ in the region where coannihilations with the lightest squark are most important. In this region, flavour violating signatures can be expected from the processes in Eqs.\ (\ref{eq:signal1}) and (\ref{eq:signal2}). They feature, however, cross sections that are smaller by about two orders of magnitude. 

For the LHC with $\sqrt{s}=14$ TeV and an integrated luminosity of 100 fb$^{-1}$, the strongest signal in Eq.\ (\ref{eq:signal0}) can in our example scenario lead to up to about $10^4$ events. As discussed in Ref.\ \cite{NMFV_Squark2}, this signature is rather clean and not subject to important backgrounds. The region, where the correct relic density is achieved through efficient coannihilation, however, forbids this particular channel. The other potentially interesting channels may then lead to about a few hundred events each. Note that the event rate, the exact dependence on the NMFV-parameters, as well as the dark matter relic density are scenario-dependent. We can, however, expect a similar behaviour for other scenarios in the NMFV-MSSM.

\section{Conclusions \label{sec6}}

While the Minimal Supersymmetric Standard Model (MSSM) with a most general flavour structure has been extensively studied in the context of collider signatures, the possibility of squark flavour mixing has not been considered for observables related to dark matter so far. However, as the LHC is running and more precise cosmological and astrophysical experiments are taking data or being set up, it will become more and more important to take into account such effects when studying the interplay between collider and astroparticle phenomenology.

In the case of neutralino dark matter in supersymmetric theories, flavour violating couplings can influence the (co)annihilation cross section, and in consequence the predicted relic density, in different ways. The strongest effect is due to the modified mass spectrum of squarks, the lightest squark becoming lighter with increasing flavour non-diagonal terms in the mass matrices. The exchange of squarks in neutralino pair annihilation as well as the presence of coannihilation with a squarks become then important. Another effect comes from the fact that couplings of neutralinos to squarks are not diagonal in flavour space any more. This opens new (co)annihilation channels, such as $\tilde{\chi}_1^0\tilde{\chi}_1^0 \to c\bar{t}$ or $\tilde{\chi}_1^0\tilde{u}_1 \to ch^0 (cg, cZ^0)$, which can give sizeable contributions to the annihilation cross-section already for moderate flavour violation parameters.

Considering flavour mixing in the sector of right-handed up-type squarks, we have shown that the modified squark masses and flavour contents have a strong impact on the (co)annihilation modes. New annihilation channels are opened due to the presence of non-diagonal couplings in flavour space. These new contributions may become numerically important in particular regions of the parameter space. As a consequence, new regions that are compatible with the relic density constraint are opened. We emphasize the fact that these new regions are not excluded by the rather strong constraints imposed by flavour physics observables. Moreover, effects of lepton flavour violation on neutralino dark matter have recently been discussed in Ref.\ \cite{LFV_CoAnn}.

A brief study of the corresponding LHC phenomenology has shown that the clean signature $pp \to c\bar{t} E_{\rm T}^{\rm miss}$, that has recently been studied in Ref.\ \cite{NMFV_Squark2}, can only be realized for moderate flavour mixing, when the lightest squark mass comes not too close to the neutralino mass. For rather large flavour violation, however, this channel is closed and NMFV-signatures arise through production and decay of gluinos rather than squarks. Such signatures include production of a top quark in association with charm-jets and may yield a few hundred events at the LHC with $\sqrt{s}=14$ TeV and an integrated luminosity of 100 fb$^{-1}$.

Since the annihilation cross section of the neutralino also governs the particle fluxes, flavour violating couplings would also have an impact on indirect detection of dark matter. In particular, additional $\tilde{c}$--$\tilde{t}$ mixing, as discussed in this paper, would change the spectrum of photons originating from dark matter annihilation. The impact of flavour mixing is, however, expected to be very small compared to the astrophysical uncertainties in this context.

Direct dark matter detection might also be influenced by the discussed flavour mixing. Here, the scattering of a neutralino off a nucleus can proceed through squark-exchange, such that the charm-content in the nucleon becomes relevant if the lightest squark is a mixture of stop and scharm. In the same way, flavour mixing in the sector of down-type squarks would increase the importance of the strange quark in the nucleus. Detailed studies of direct or indirect detection of dark matter in the context of flavour violation are, however, beyond the scope of this work.

\begin{acknowledgments}
The authors would like to thank W.~Porod, A.~Pukhov and F.~Staub for their help concerning the used computer programs and S.~Kraml for helpful discussions. This work is supported by Helmholtz Alliance for Astroparticle Physics and by the Landes-Exzellenzinitiative Hamburg. The work of Q.L.B.\ is supported by a Ph.D.\ grant of the French Ministry for Education and Research.
\end{acknowledgments}


\end{document}